\documentclass[12pt]{article}

\usepackage{scicite}

\usepackage{times}

\usepackage[labelfont=bf,labelsep=period]{caption}

\usepackage{amsmath,mathtools}
\usepackage{hyperref,graphicx,multirow,multicol,color}
\usepackage[version=4]{mhchem}
\usepackage{booktabs}
\usepackage{lscape}
\usepackage{longtable}
\usepackage{rotating}
\usepackage{adjustbox}
\usepackage{comment}

\usepackage[utf8]{inputenc}

\newenvironment{sciabstract}{%
\begin{quote} \bf}
{\end{quote}}

\newcommand{\sifigurename}{Supplementary Figure}
\newcommand{\sitablename}{Supplementary Table}
\newcommand{\sifigurenames}{Supplementary Figures}
\newcommand{\sitablenames}{Supplementary Tables}

\newcounter{lastnote}

\title{ Ubiquitous Aromatic Carbon Chemistry at the Earliest Stages of Star Formation}

\author {Andrew M. Burkhardt,$^{1,\ast}$ Ryan A. Loomis,$^{2}$\\  Christopher N. Shingledecker,$^{3,4,5}$ Kin Long Kelvin Lee,$^{6,1}$ Anthony J. Remijan,$^2$\\ Michael C. McCarthy,$^{1}$ and Brett A. McGuire$^{6,1,2,\ast}$ \\
\\
\normalsize{$^{1}$Center for Astrophysics $\mid$ Harvard~\&~Smithsonian, Cambridge, MA 02138, USA}\\
\normalsize{$^{2}$National Radio Astronomy Observatory, Charlottesville, VA 22903, USA}\\
\normalsize{$^{3}$Department of Physics and Astronomy, Benedictine College, Atchison, KS 66002, USA}\\
\normalsize{$^{4}$Center for Astrochemical Studies, Max Planck Institute for}\\ 
\normalsize{Extraterrestrial Physics, Garching, Germany}\\
\normalsize{$^{5}$Institute for Theoretical Chemistry, University of Stuttgart, Stuttgart, Germany}\\
\normalsize{$^{6}$Department of Chemistry, Massachusetts Institute of Technology, Cambridge, MA 02139, USA}\\
\normalsize{$^\ast$To whom correspondence should be addressed;}\\
\normalsize{E-mails: andrew.burkhardt@cfa.harvard.edu, brettmc@mit.edu.}
}

\date{}

\begin{document} 


\baselineskip24pt


\maketitle


\begin{sciabstract}
 Benzonitrile ($c$-\ce{C6H5CN}), a polar proxy for benzene ($c$-\ce{C6H6}), has the potential to serve as a highly convenient radio probe for aromatic chemistry, provided this ring can be found in other astronomical sources beyond the molecule-rich prestellar cloud TMC-1 where it was first reported by McGuire \textit{et al.} in 2018. Here we present radio astronomical evidence of benzonitrile in four additional pre-stellar, and possibly protostellar, sources: Serpens 1A, Serpens 1B, Serpens 2, and MC27/L1521F. These detections establish benzonitrile is not unique to TMC-1; rather aromatic chemistry appears to be widespread throughout the earliest stages of star formation, likely persisting at least to the initial formation of a protostar. The abundance of benzonitrile far exceeds predictions from models which well reproduce the abundances of carbon chains, such as \ce{HC7N}, a cyanpolyyne with the same heavy atoms, indicating the chemistry responsible for planar carbon structures (as opposed to linear ones) in primordial sources is favorable but not well understood. The abundance of benzonitrile relative to carbon-chain molecules displays sizable variations between sources within the Taurus and Serpens clouds, implying the importance of physical conditions and initial elemental reservoirs of the clouds themselves.

\end{sciabstract}

Aromaticity has long been believed to be an integral aspect of astrochemistry. The Unidentified Infrared Bands (UIRs), a set of emission features observed at mid-IR wavelengths (roughly from 3 to 13 $\mu$m), are thought to originate from the vibrational relaxation of polycyclic aromatic hydrocarbons (PAHs) \cite{Leger:1984dd, Allamandola:1985vw, Tielens:2008fx} following electronic excitation. These bands have been observed in an astonishing range of astrophysical environments, from the expanding atmospheres of Asymptotic Giant Branch (AGB) stars or supernovae \cite{Gauhub:2004}, to the latest stages of star formation \cite{Bregman:2001}, to the gas in external galaxies \cite{2007ApJ...656..770S}. Based on these observations it has been estimated that as much as 10-25\% of interstellar carbon may be locked up in aromatic molecules \cite{Tielens:2008fx}, making these species important in the chemistry in nearly all regions. In addition, the nucleation and accretion of the largest of these molecules may be a key driver for the growth of carbonaceous interstellar dust grains in the atmosphere of certain evolved stars. Among the more than 200 molecules definitively detected in the interstellar medium to date \cite{McGuire:2018mc}, however, only a very small fraction are aromatic either in a strictly chemical sense or otherwise: cyclopropenylidene ($c$-\ce{C3H2}), benzene (\ce{C6H6}), three fullerenes (\ce{C60}, \ce{C60^+}, \ce{C70}) \cite{Thaddeus:1985hw,Vrtilek:1987iu,Foing:1994,Foing:1997,Cernicharo:2001mw,Cami:2010fi,Berne:2013ow}, and the simplest aromatic nitrile benzonitrile \ce{C6H5CN}, which was recently identified in the molecule-rich, prestellar Taurus Molecular Cloud (TMC-1) by radio astronomy \cite{McGuire:2018it}. Despite these discoveries, very little is known about how these specific aromatic molecules fit within the broader context of interstellar chemistry. 

Based on extensive laboratory and theoretical work, it is now well established that the formation of benzonitrile from benzene and a source of chemically active nitrogen, commonly CN radical, is efficient under interstellar conditions \cite{Balucani:1999it,trevitt:1749,Lee:2019dh, McGuire:2018it,Cooke:2020we}. The presence of benzonitrile is thus a direct and meaningful indicator of radio-invisible benzene. Although the column densities of many carbon-chains in cold cores such as TMC-1 are well reproduced by astrochemical models based on reactions starting with small hydrocarbon precursors \cite{Cordiner:2012fk,Loomis:2016js,Burkhardt:2018ka,shingledecker_cosmic-ray-driven_2018}, that of benzonitrile is in excess of predictions from the same model, suggesting other routes to form aromatic species may be operative in this source \cite{McGuire:2018it}.

In this context, two consequential lines of inquiry in dark cloud chemistry follow: the extent to which the unusually-rich chemistry in TMC-1 is broadly representative of dark clouds, and how differences in cloud properties affect this chemistry. TMC-1 is known to harbor unsaturated carbon-chains in remarkably high abundance even relative to other dark clouds \cite{Kaifu:2004tk,Gratier:2016fj}. Many such chains were first or only detected in the interstellar medium toward this source \cite[and references therein]{McGuire:2018mc}. However, whether the chemical inventories of dark clouds commonly extend to aromatic molecules has been a topic largely confined to conjecture until very recently. The newfound ability to infer the population of benzene by radio astronomy now affords one the opportunity to assess cloud chemistry from a distinctly different viewpoint. This fundamental organic ring, owing to its much higher degree of saturation, may be formed by pathways quite unlike those invoked for highly unsaturated carbon-chains, and is thought to be the key building block in the synthesis of far more complex organic molecules \cite{Cherchneff:1992, Phillips:1999hd, Kim:2011es}. To substantively advance this discussion, however, it is first necessary to establish if benzonitrile, and thus benzene by inference, is present beyond TMC-1 and, if so, to what degree this ring is widespread in molecular clouds.

Long carbon-chain molecules, such as the well-known cyanopolyynes (HC$_n$N, where $n$ is odd), are readily observed throughout the various stages of star formation, and are particularly common at the earliest stages, prior to the desorption of saturated molecules from ice which occurs when the protostellar object warms up the surrounding material \cite{Garrod:2013id}. And, in particular, high column densities of large cyanopolyynes, such as \ce{HC7N} and \ce{HC9N}, can provide evidence for a rich source of complex hydrocarbons. Given \ce{HC7N}, a highly unsaturated (H$\ll$C) chain, has the same heavy atoms as benzonitrile, we searched for radio lines of benzonitrile toward sources with high column densities of this species as an indicator of a rich gas-phase carbon chemistry with many of the essential building blocks to produce benzene. We discuss four sources with $N_{\ce{HC7N}} > 10^{12}$ cm$^{-2}$ \cite{Sakai:2010bk,Friesen:2013ii} that were first targeted: Serpens 1a (S1a), Serpens 1b (S1b), Serpens 2 (S2), and MC27/L1521F, along with TMC-1, where benzonitrile was originally discovered.

The five sources reside in two distinct parent clouds and probe different stages of prestellar environments. The Serpens sources are all part of the same greater filament studied by Friesen et al. \cite{Friesen:2013ii}. Based on its molecular line widths, lack of embedded protostars, and small continuum-derived mass, S2 is thought to be a very young starless core, with a maximum estimated age of 10$^5$ years \cite{Friesen:2013ii}. S1a and S1b are two cores straddling a central protocluster where infall is occurring and have estimated ages of several 10$^5$ years \cite{Friesen:2013ii}. Both TMC-1 and MC27/L1521F are associated with the larger Taurus Molecular Cloud Structure. TMC-1 is a well-studied prestellar molecular cloud with minimal infall motion, whose age has been estimated to be $\sim$2-5$\times$10$^5$ years \cite{hincelin_oxygen_2011,Majumdar:2016dn,McGuire:2017ud, McGuire:2018it,Loomis:2020aa}. MC27/L1521F is considered a Very Low Luminosity Object (VeLLO), where the embedded protostar is just beginning to heat up \cite{Bourke:2006kn}.

Our observations were performed with the 100\,m Robert C. Byrd Green Bank Telescope (GBT) as part of a large ongoing observing campaign: A Rigorous K/Ka-Band Hunt for Aromatic Molecules (ARKHAM), which is slated to survey more than dozen sources when it is complete. An updated analysis of \ce{C6H5CN} in TMC-1 \cite{Loomis:2020aa} has also been possible with additional observations from our companion survey, GOTHAM (GBT Observations of TMC-1: Hunting Aromatic Molecules; \cite{McGuire:2020bb}). For each source, four strong benzonitrile transitions were targeted between 22 and 26.5\,GHz with the K-band Focal Plane Array \cite{Morgan:2008kb} and VEGAS spectrometer \cite{Roshi:2012he} (\sitablenames~ \ref{tab:observations_table} and \ref{tab:spectra_table}). For S2 and MC27/L1521F, additional observations were performed with the Ka-band receiver and target four other transitions between 28 and 30.5\,GHz. Additionally, MC27/L1521F also has several lower frequencies across K-band between 22 and 26\,GHz. The GOTHAM observations towards TMC-1 cover transitions in X, K, and Ka-band, and are discussed in detail elsewhere \cite{McGuire:2018it,McGuire:2020bb}.

\begin{figure}
    \centering
    \vspace{-2cm}\includegraphics[width=0.34\textwidth]{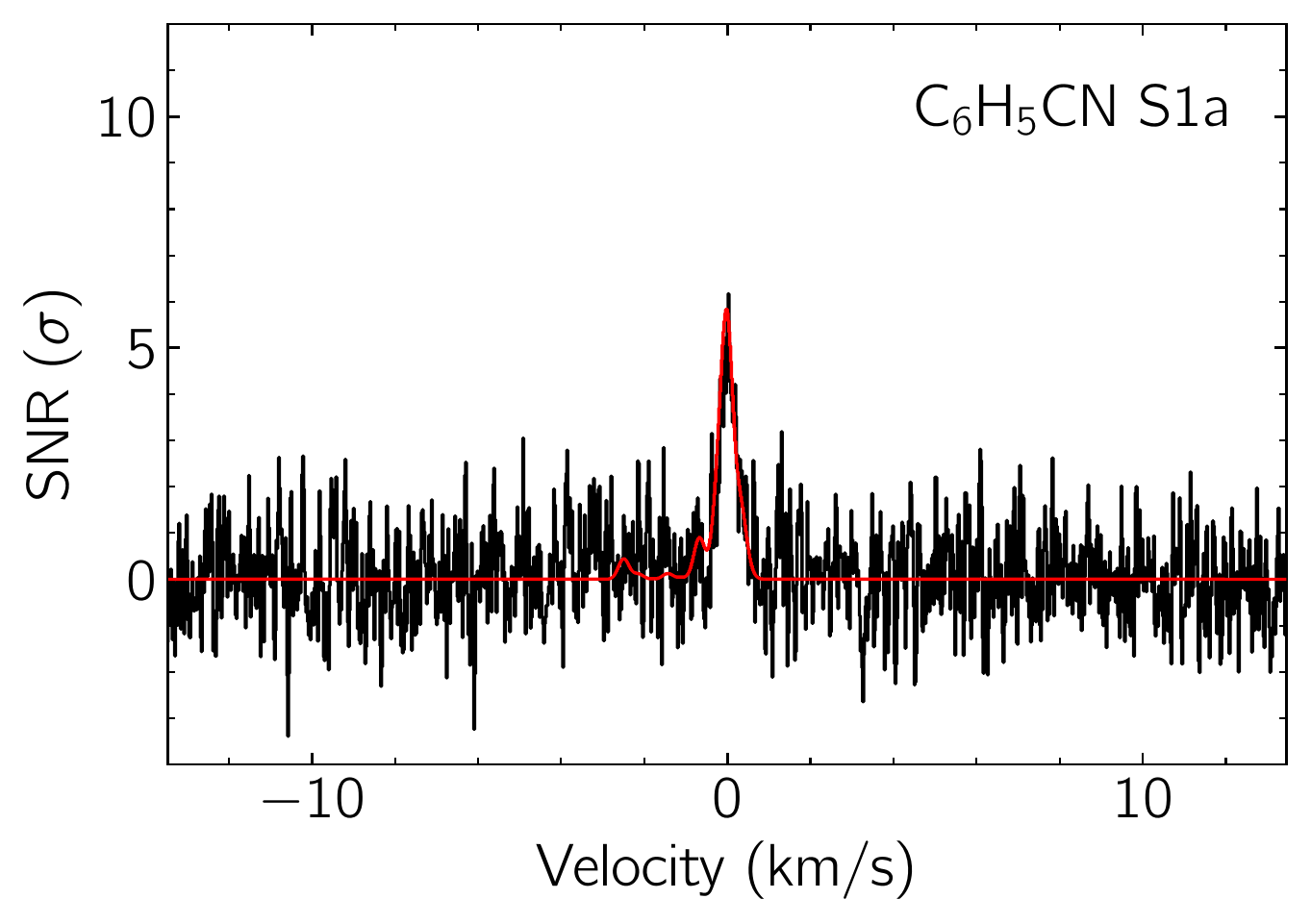}
    \includegraphics[width=0.34\textwidth]{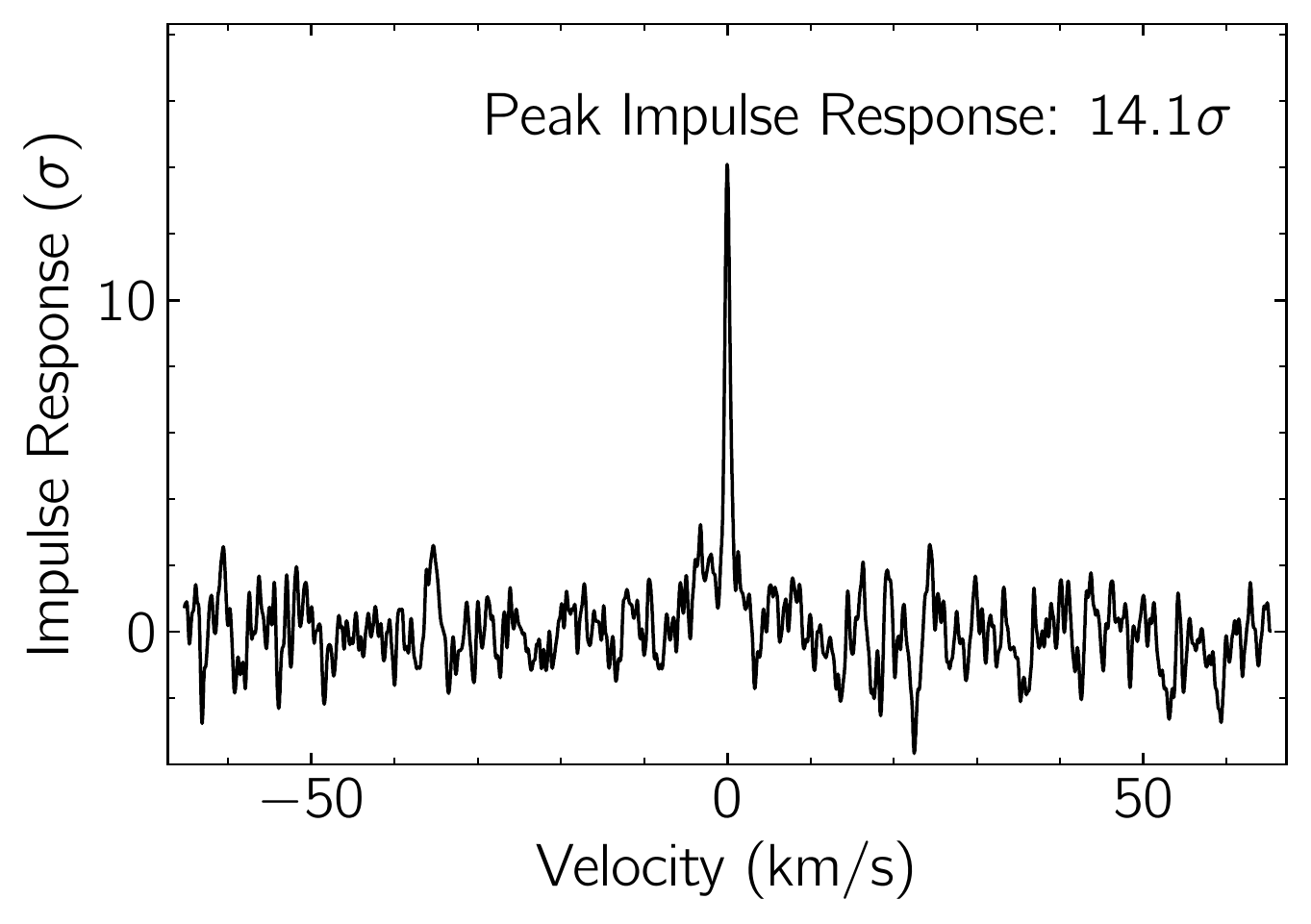}
    \includegraphics[width=0.34\textwidth]{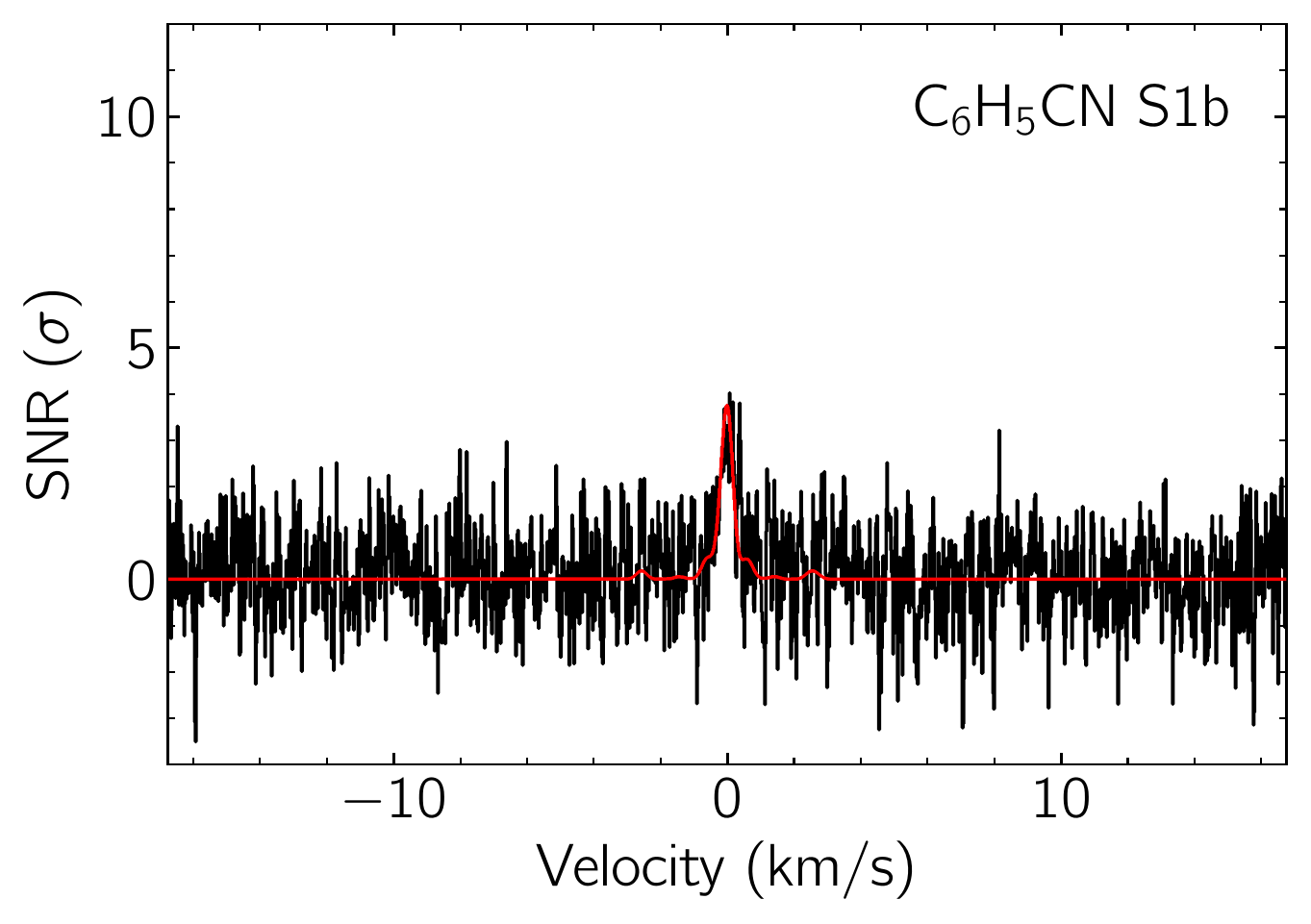}
    \includegraphics[width=0.34\textwidth]{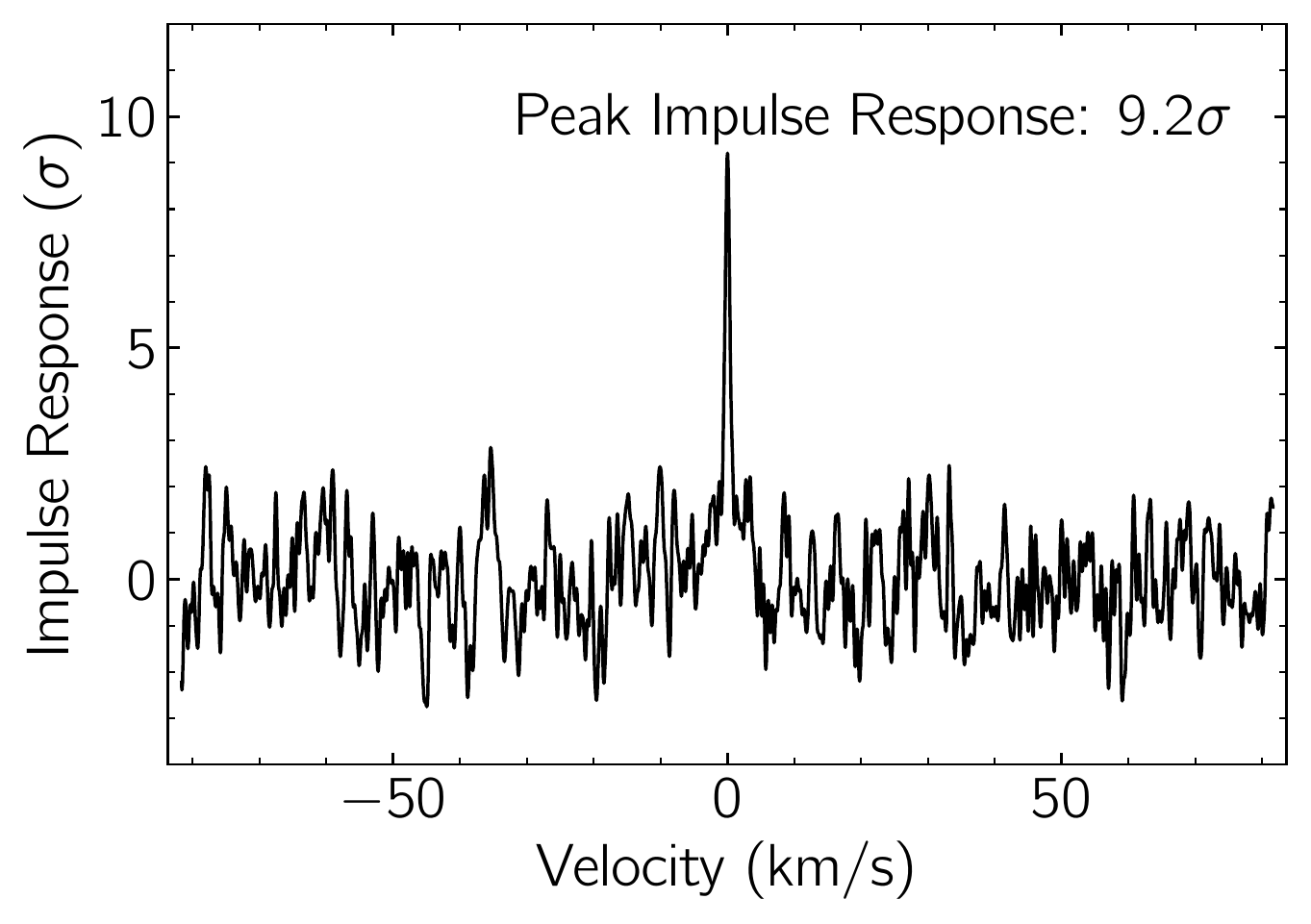}
    \includegraphics[width=0.34\textwidth]{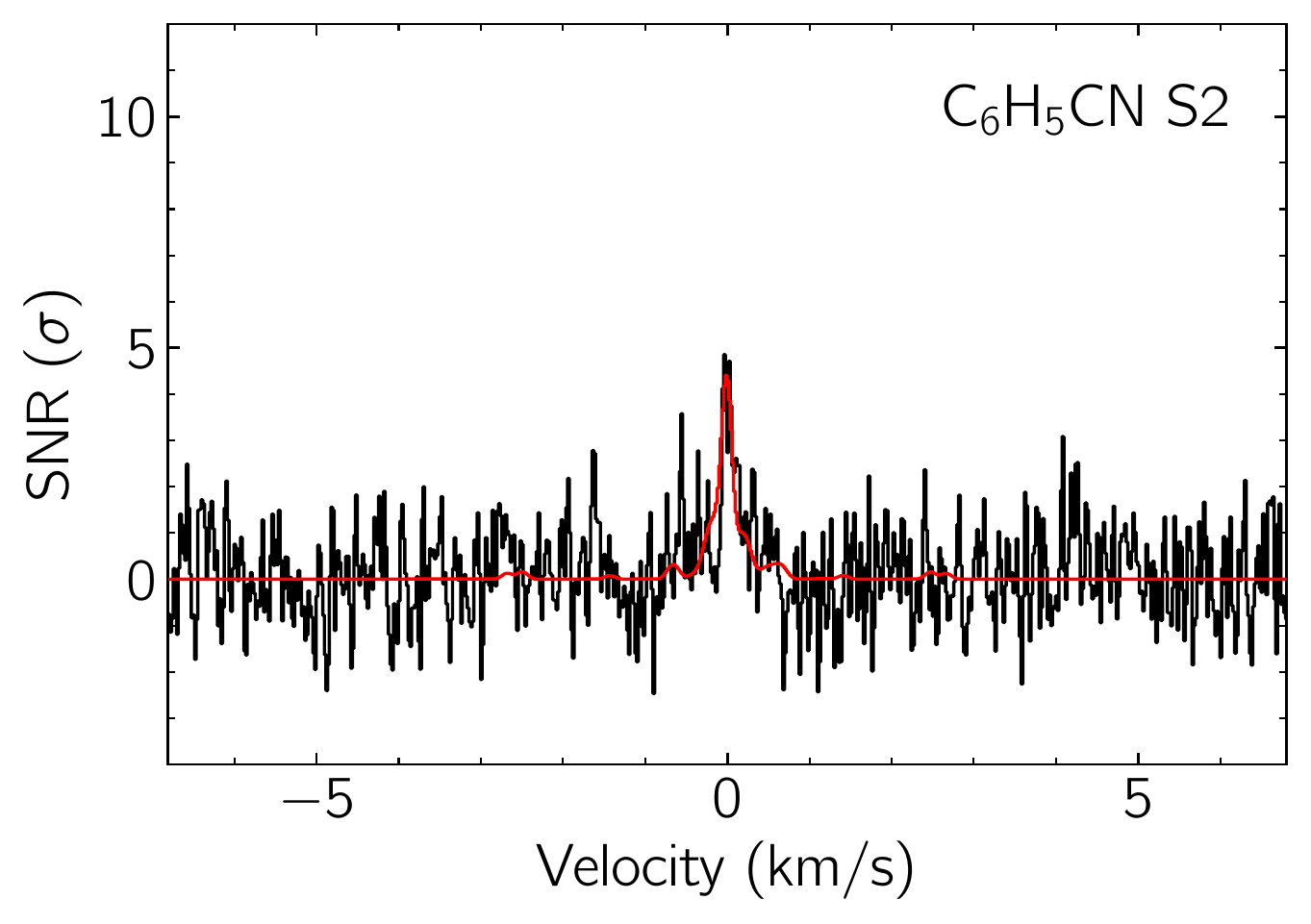}
    \includegraphics[width=0.34\textwidth]{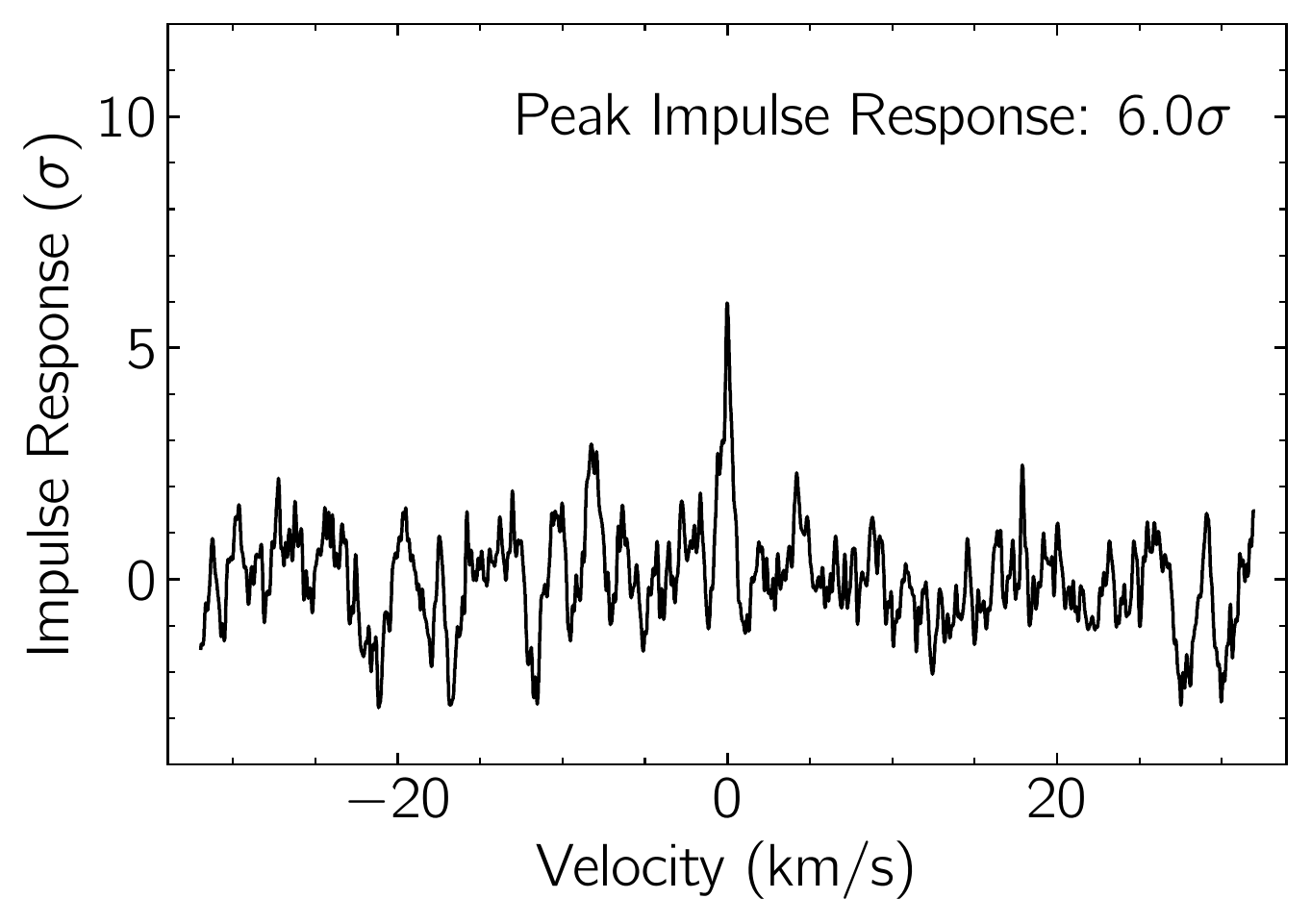}
    \includegraphics[width=0.34\textwidth]{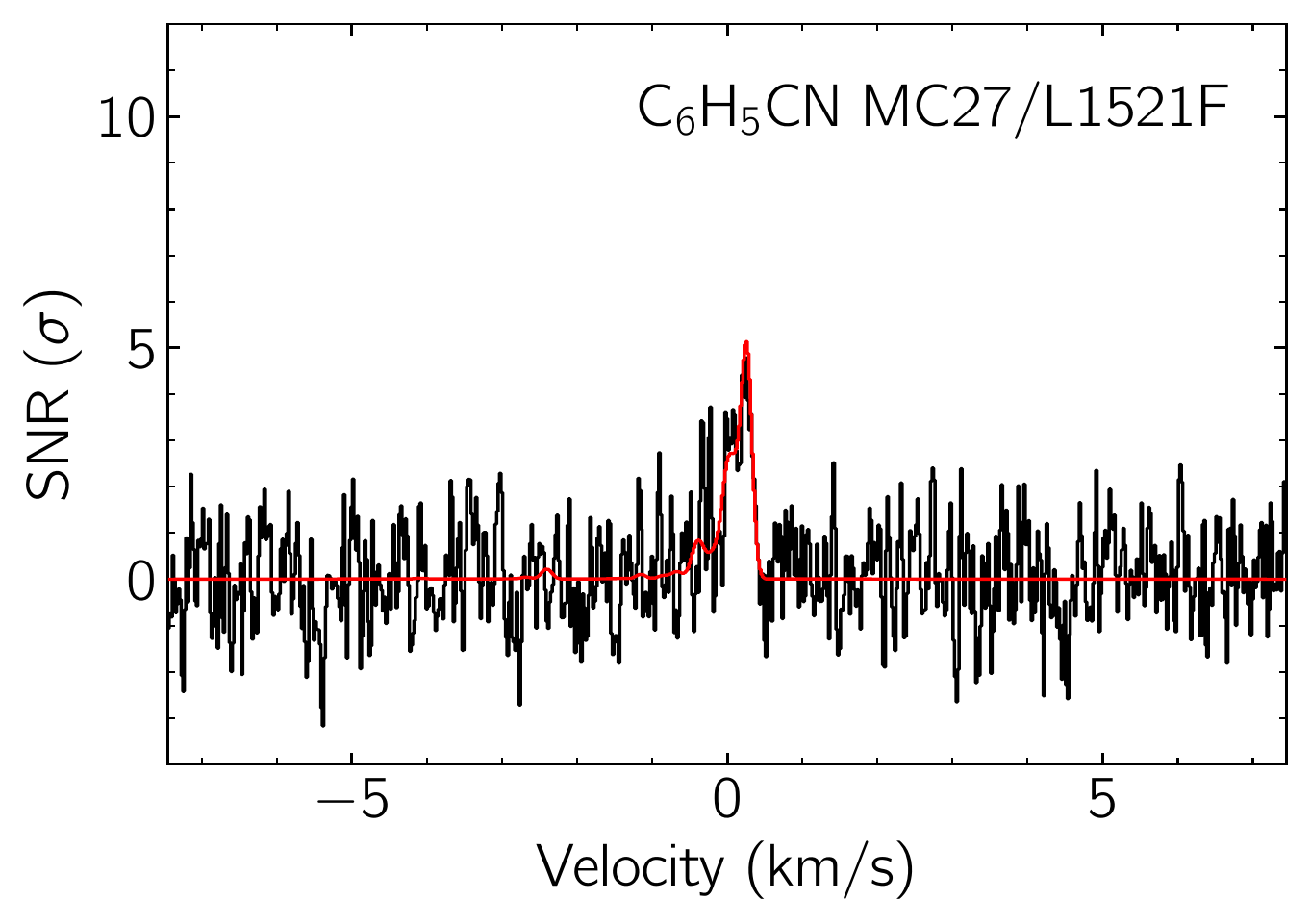}
    \includegraphics[width=0.34\textwidth]{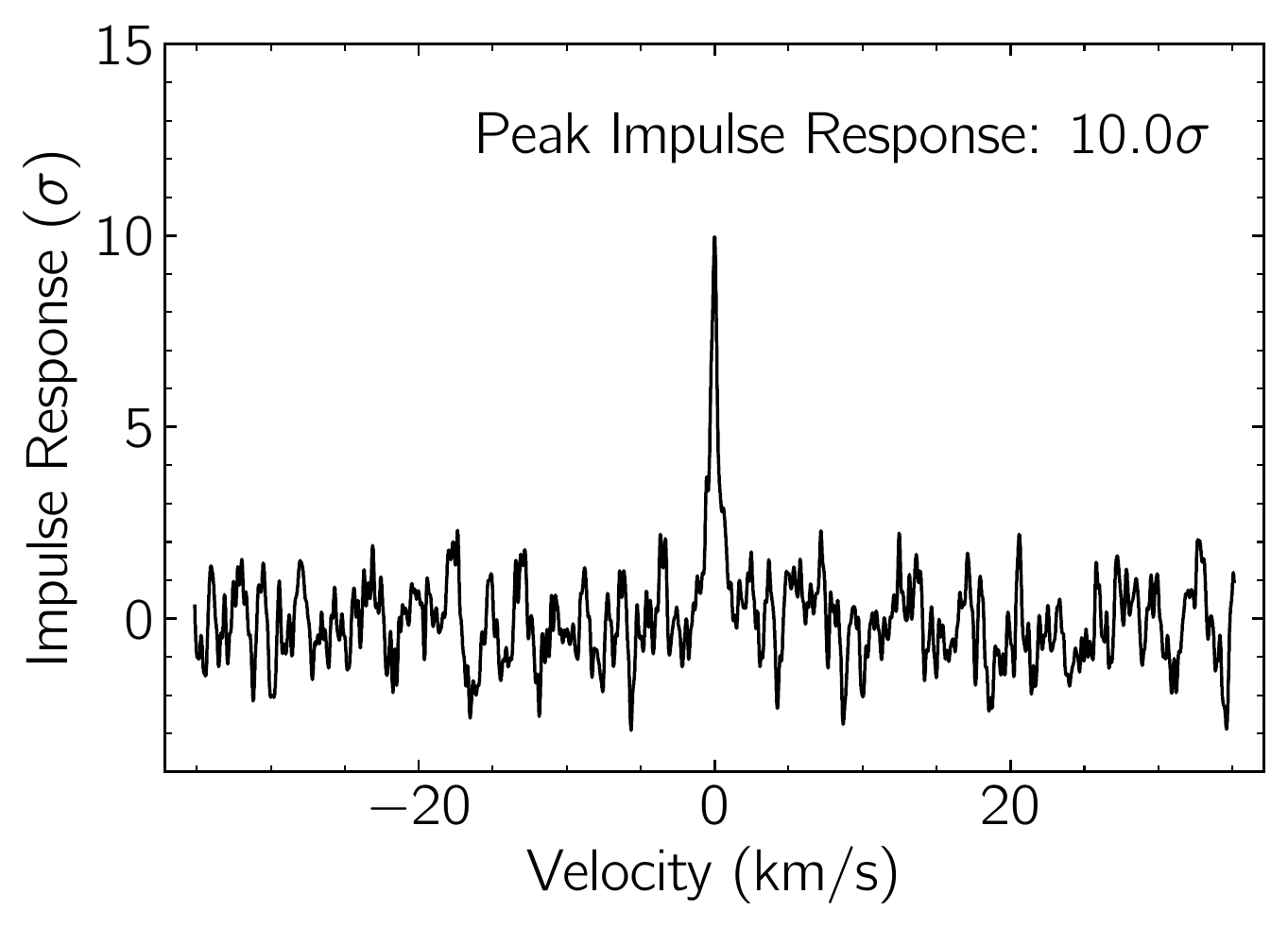}
    \includegraphics[width=0.34\textwidth]{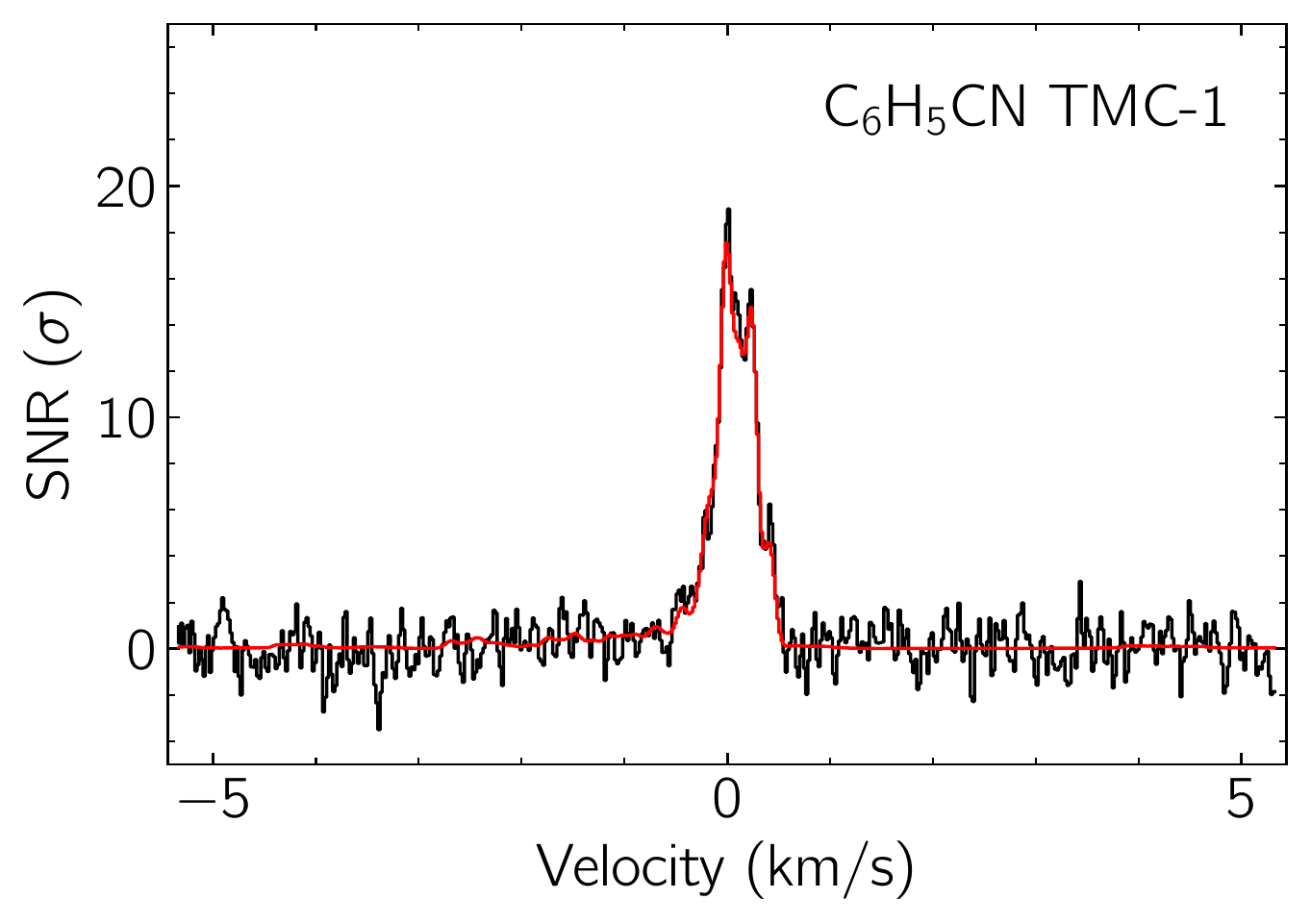}
    \includegraphics[width=0.34\textwidth]{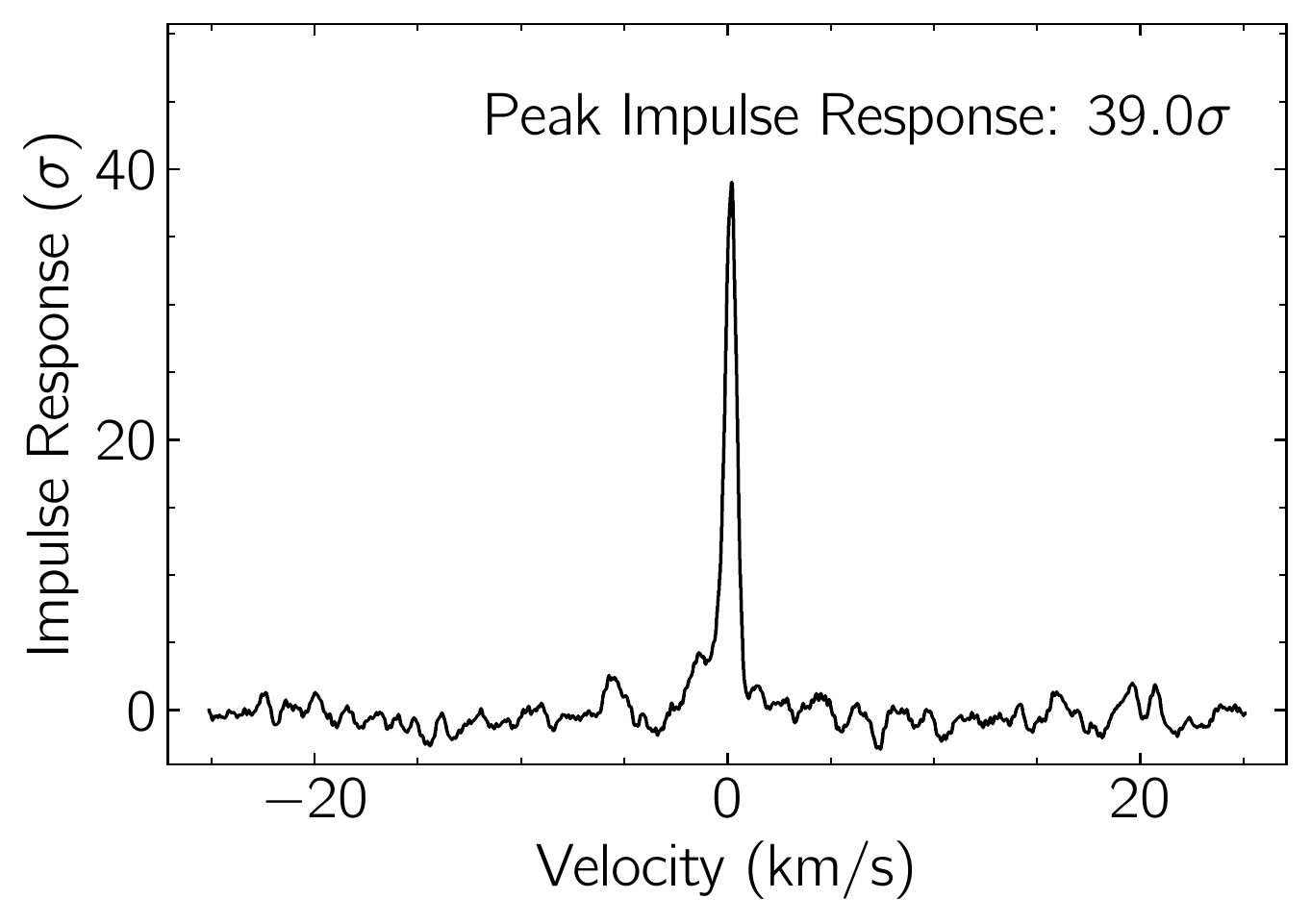}
    \caption{Velocity-stacked spectra of \ce{C6H5CN} and the impulse response function of the stacked spectra in the five sources observed here. \textit{Left panel}: Velocity-stacked spectra are in black, with the corresponding stack of the simulation using the best-fit parameters to the individual lines in red. The data have been uniformly sampled to a resolution of 0.02\,km\,s$^{-1}$. The intensity scale is the signal-to-noise ratio of the spectrum at any given velocity. \textit{Right panel}: The impulse response function of the stacked spectrum generated using the simulated line profile as a matched filter. The intensity scale is the signal-to-noise ratio of the response function when centered at a given velocity. The peak of the impulse response function provides a minimum significance for the detection noted in each figure. See Loomis et al.\cite{Loomis:2020aa} for details.}
    \label{fig:bn_stack}
\end{figure}

Three transitions of benzonitrile were ultimately detected in each of our new four target sources (\sifigurenames~\ref{s2_bn_spectra}, \ref{s1a_bn_spectra}, \ref{s1b_bn_spectra}, and \ref{mc27_bn_spectra}). The same analysis procedure for the GOTHAM data \cite{Loomis:2020aa} was employed here, in which a Markov Chain Monte Carlo (MCMC) fit is used to determine the column density ($N_T$), excitation temperature ($T_{ex}$), linewidth ($\Delta V$), and source velocity ($v_{lsr}$) or velocities that best reproduce the observations; details of this procedure are given in the Spectral Stacking Routine section of Methods, with a complete description provided in Ref.~\cite{Loomis:2020aa}. Transition frequencies, line strengths, and the partition function for \ce{C6H5CN} were taken from previous laboratory measurements \cite{wohlfart:119,McGuire:2018it}. The velocity-stacked spectra from transitions toward each of the five sources (Fig.~\ref{fig:bn_stack}) show that benzonitrile is detected at least 5$\sigma$ in both the velocity-stacked spectra and peak impulse response of the stack. The relative ease with which this polar ring has been observed in all five sources suggests that aromatics can survive through the formation of a protostar and perhaps beyond.

\begin{figure}
    \centering
    \includegraphics[width=\textwidth]{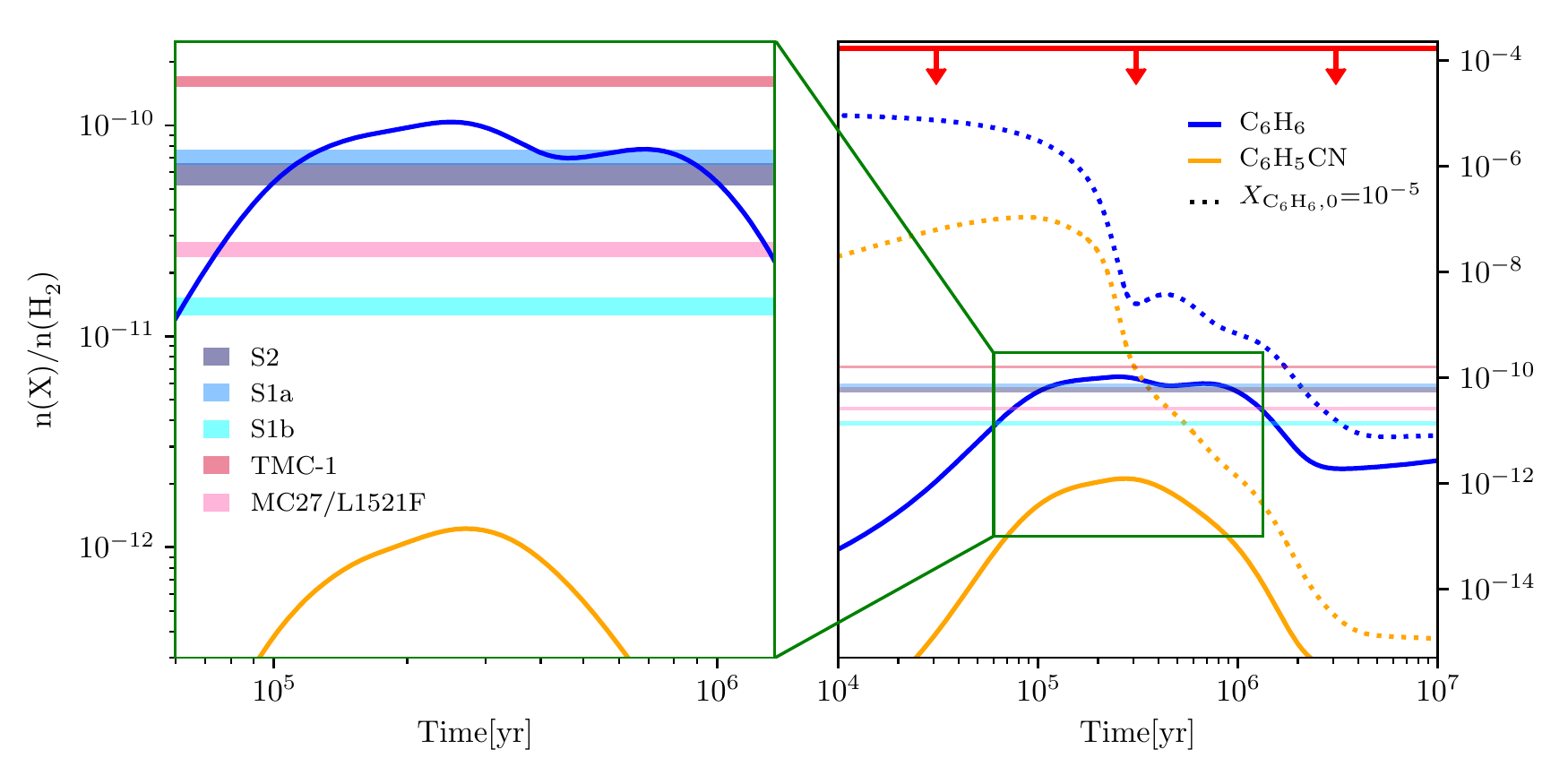} 
    \caption{Simulated abundances from \textsc{nautilus} chemical models for \ce{C6H6} and \ce{C6H5CN} (solid) in comparison to that derived for \ce{C6H5CN} from observations in the five sources studied here, with uncertainties (as described in \sitablenames~\ref{s2_bn_results}, \ref{s1a_bn_results}, \ref{s1b_bn_results}, \ref{mc27_bn_results} and \ref{mc27_bn_results} and \ref{tmc1_bn_results}) shown in horizontal bars in the same colors as Fig. \ref{fig:ratios}.
    \emph{Left:} Zoom in on time frame relevant for ARKHAM sources, highlighting the observed abundances are far greater than what existing models predict. \emph{Right:} Zoom out of \emph{Left}, in addition to simulation with an initial \ce{C6H6} abundance of 10$^{-5}$ (dotted). The initial elemental carbon abundance, or the model's total reactive carbon budget, is shown in red with downward arrows.}
    \label{fig:obs_and_model}
\end{figure}

Although spectra from all five sources were fit with one or more velocity components, the present single-dish data provide little meaningful spatial information. As such, we first examined the total abundance of \ce{C6H5CN} in each source across all velocity components. For all but TMC-1, the primary uncertainty in the derived column densities arises from the degeneracy between this quantity and the excitation temperature, which can be mitigated by observing transitions over a still wider range of upper level energies. Toward TMC-1, the largest uncertainty is due to the unconstrained source sizes of the velocity components. Ultimately, high-sensitivity mapping with the Very Large Array in a compact configuration or ARGUS mapping with GBT is needed to better constrain the spatial extent of the emission features. Nevertheless, as indicated in Fig.~\ref{fig:obs_and_model}, the \ce{C6H5CN} abundance relative to \ce{H2} ranges between approximately 10$^{-11}$ and 10$^{-10}$, with the highest values toward TMC-1. The Supplementary Information provide significant additional information on the uncertainties, source sizes, derived column densities, etc.~for each source studied here.

Overlaid on Fig.~\ref{fig:obs_and_model} are abundance predictions of \ce{C6H5CN} and \ce{C6H6} from a gas-grain kinetic chemical model, \textsc{nautilus} \cite{Ruaud:2016bv}, which has been extended beyond that reported in Ref.~\cite{McGuire:2018it} and \cite{shingledecker_cosmic-ray-driven_2018} to account for new molecular identifications in the GOTHAM and ARKHAM surveys [see the Astrochemical Modeling section of Methods; Ref.~\cite{Loomis:2020aa,McGuire:2020aa,McGuire:2020bb,McCarthy:2020aa,Xue:2020aa}]. It is clear from Fig.~\ref{fig:obs_and_model} that the derived abundances of benzonitrile are consistently at least an order of magnitude higher than those predicted from our most up-to-date models, implying aromaticity is more important in prestellar chemistry than previously thought. In contrast, the same model well reproduces the abundance of \ce{HC7N} and even \ce{HC9N}~\cite{Loomis:2020aa}. Taken together, we conclude the formation of aromatic molecules is quite favorable but significantly less constrained than that of highly unsaturated carbon chains.

With respect to TMC-1 specifically, we note the observationally derived and model-predicted abundances of benzonitrile differ from those reported earlier \cite{McGuire:2018it}. In terms of the derived abundance, the more accurate methods adopted here and in the GOTHAM survey \cite{Loomis:2020aa} result in a four times higher value. Regarding the chemical models, the predicted abundances are lower by about an order of magnitude primarily for two reasons. First, the use of slightly different initial elemental abundances in our latest model \cite{Loomis:2020aa} produce cyanopolyynes in higher abundance, but are less efficient for cyclic species. And second, the addition of new reactions [see \textit{Supplementary Information}, Ref.~\cite{McGuire:2020aa}, and Ref.~\cite{McCarthy:2020aa}] in our network to produce other newly discovered aromatic molecules reduce the abundance of benzene and by extension benzonitrile.

The underproduction of cyclic molecules in our models is not limited to benzonitrile but extends to three other ring species detected with GOTHAM: 1- and 2-cyanonaphthalene \cite{McGuire:2020aa}, a pair of CN-functionalized PAHs (\ce{C10H7CN}), and cyano-cyclopentadiene \cite{McCarthy:2020aa}, a highly polar five-membered ring ($c$-\ce{C5H5CN}). For both rings, deviations between derived and predicted abundances are even more disparate than for benzonitrile. Nevertheless, this same model well reproduces the abundance of two newly-discovered nitrile-terminated chains, propargyl cyanide (\ce{HCCCH2CN}) \cite{McGuire:2020bb} and \ce{HC11N} \cite{Loomis:2020aa}, that are described in accompanying GOTHAM papers. This finding again suggests that our present understanding of interstellar aromatic chemistry is demonstratively incomplete.

\begin{figure}[!h]
\centering
\includegraphics[width=0.95\textwidth]{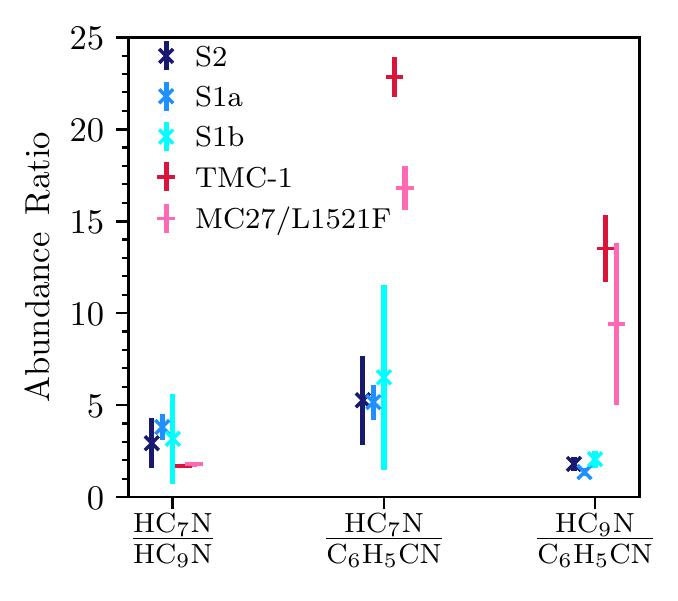}
\caption{Derived abundance ratios between \ce{HC7N}, \ce{HC9N}, and \ce{C6H5CN} for each of the five sources studied here. The colors and markers are based on the parent cloud in which they reside: blue crosses for Serpens and red dashes for Taraus. The \ce{HC7N} abundances for S2 and S1b are from a previous GBT Survey by Friesen et al. \cite{Friesen:2013ii}. Uncertainties are described in \sitablenames~\ref{s2_bn_results}, \ref{s1a_bn_results}, \ref{s1b_bn_results}, \ref{mc27_bn_results}, and \ref{tmc1_bn_results} (last of which is adapted from Table 2 in Ref~\cite{McGuire:2020bb})}
\label{fig:ratios}
\end{figure}

To determine how rich aromatic chemistry is outside of the carbon-chain rich source of TMC-1, we compared the relative abundance ratios between \ce{HC7N}, \ce{HC9N}, and \ce{C6H5CN} in each source (Fig.~\ref{fig:ratios}). For S1b and S2, the column densities of \ce{HC7N} were adopted from \cite{Friesen:2013ii}, whose observations were also done with the KFPA on the GBT. While the inverse relation of cyanopolyyne abundances with chain length \cite{Remijan:2006yd,Loomis:2016js} is fairly consistent between all five sources, the abundance ratios involving \ce{C6H5CN} vary considerably between the Taurus and Serpens sources. In each of the three Serpens sources, the \ce{HC7N}/\ce{HC9N} ratio is comparable to the chain/ring ratio using the \ce{HC7N}/\ce{C6H5CN} and \ce{HC9N}/\ce{C6H5CN} ratios as metrics. In contrast, in TMC-1 and to a lesser extent MC27, the \ce{HC7N}/\ce{C6H5CN} ratio is considerably higher, indicating aromatic chemistry is less prevalent relative to carbon-chain chemistry among sources in Taurus. We note the derived excitation temperatures are also quite similar among sources from the same parent cloud, $T_{ex}\sim$ 11.1-11.5\,K for the Serpens sources and 4.9-6.1\,K for the two sources in Taurus (MC27/L1521F and TMC-1), notwithstanding the small sample size and aforementioned degeneracy between the excitation temperatures and the derived column densities. 

\begin{figure}
    \centering
    \includegraphics[width=0.49\textwidth]{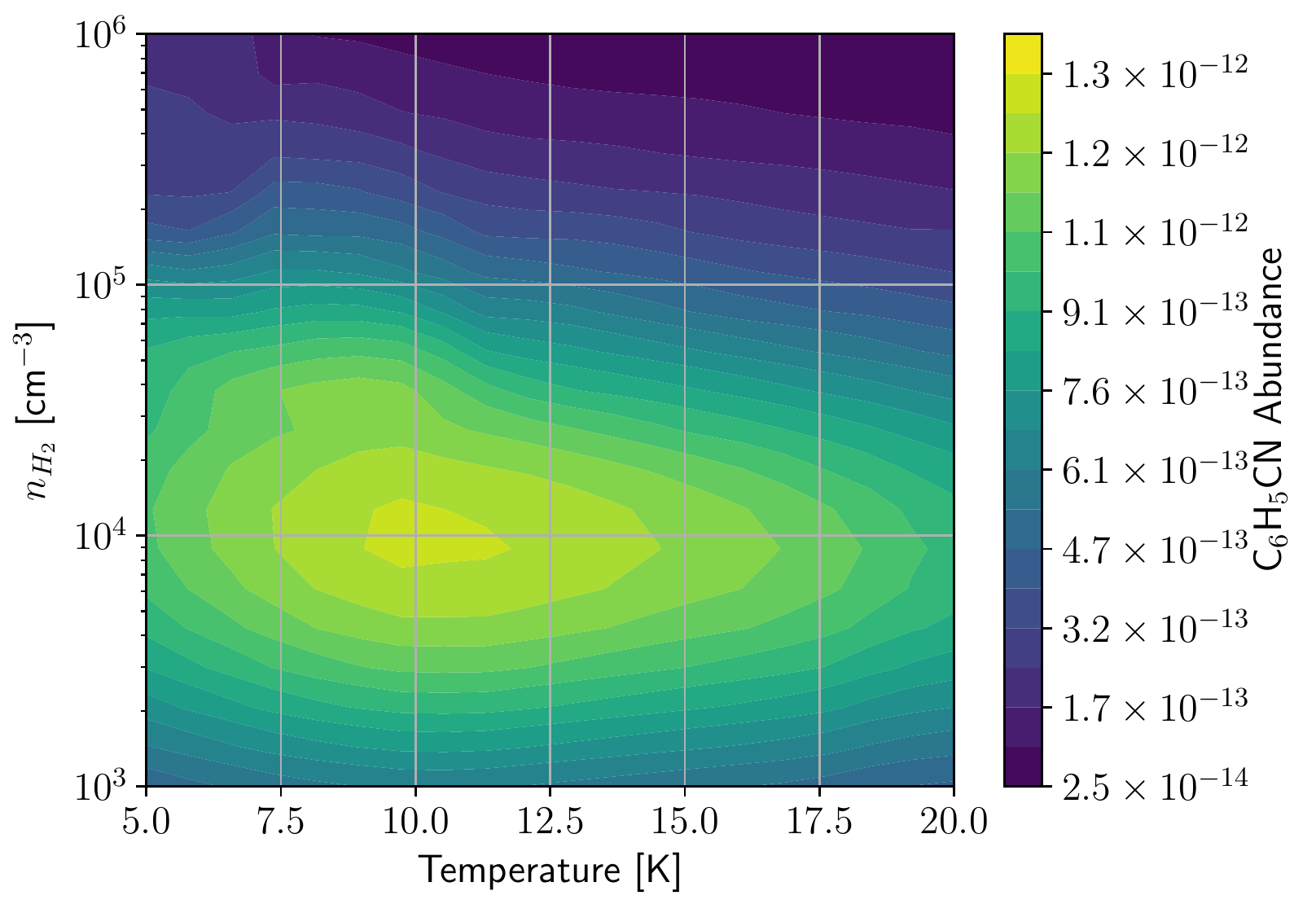}
    \includegraphics[width=0.49\textwidth]{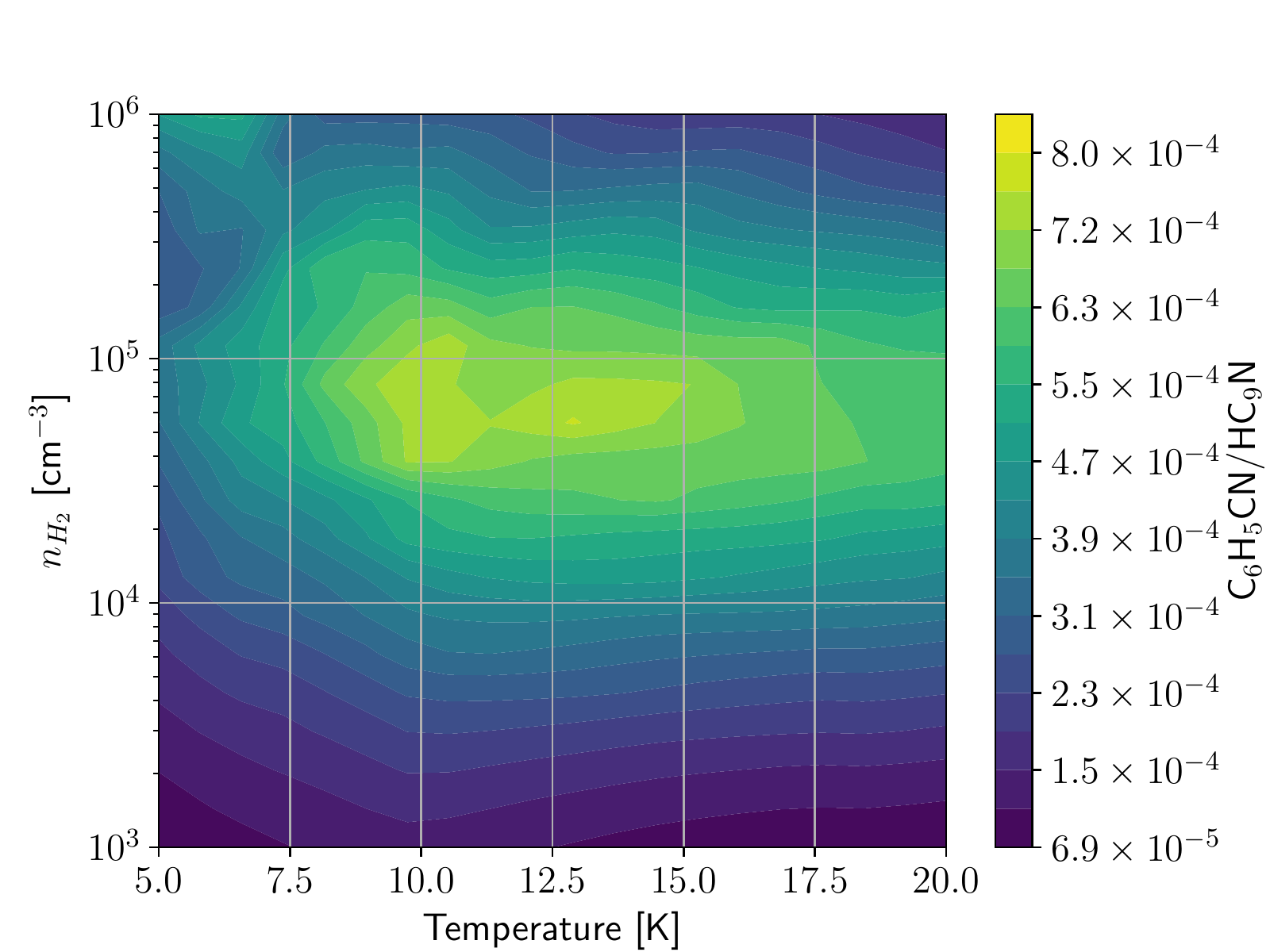}
    \includegraphics[width=0.49\textwidth]{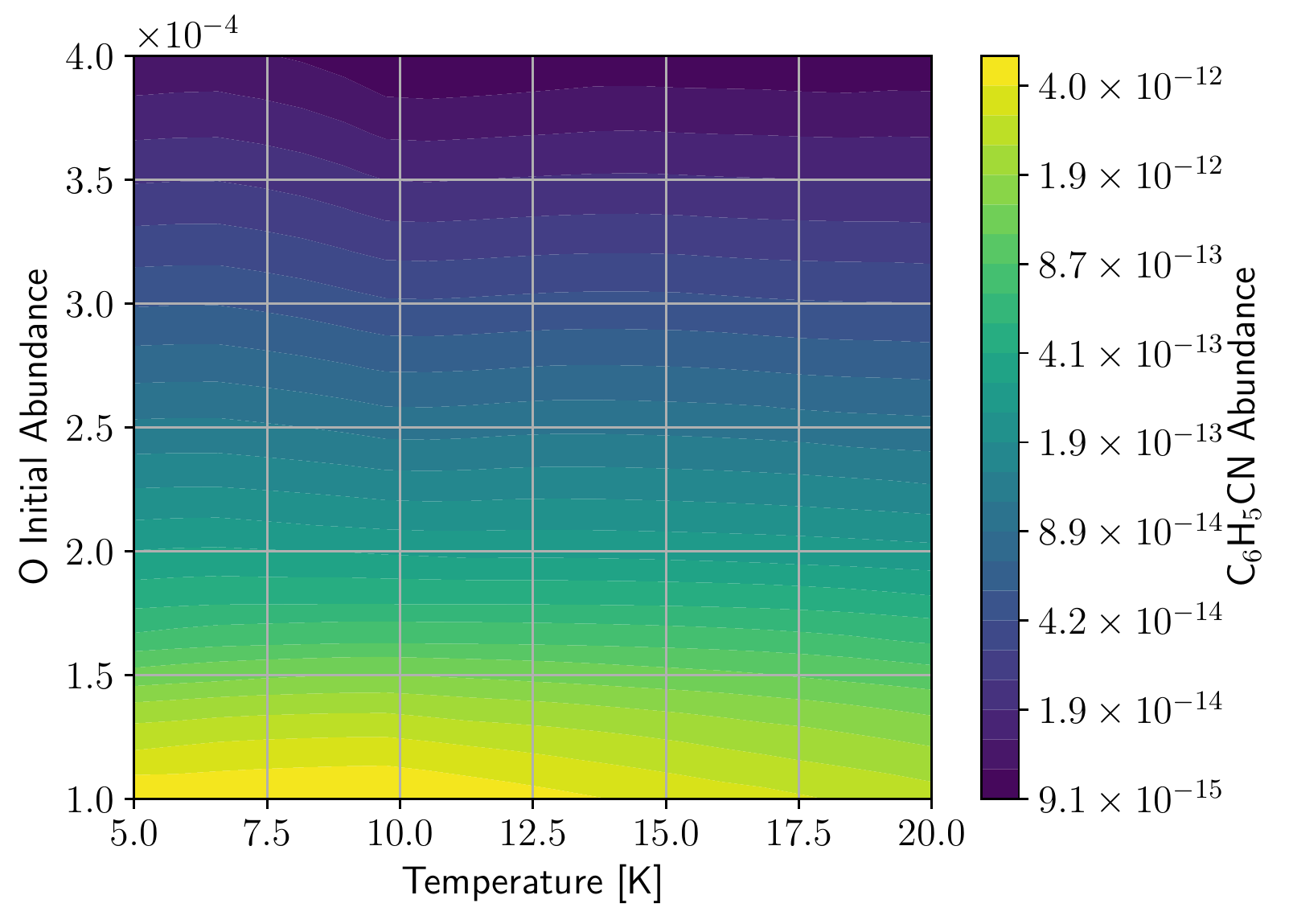}
    \includegraphics[width=0.49\textwidth]{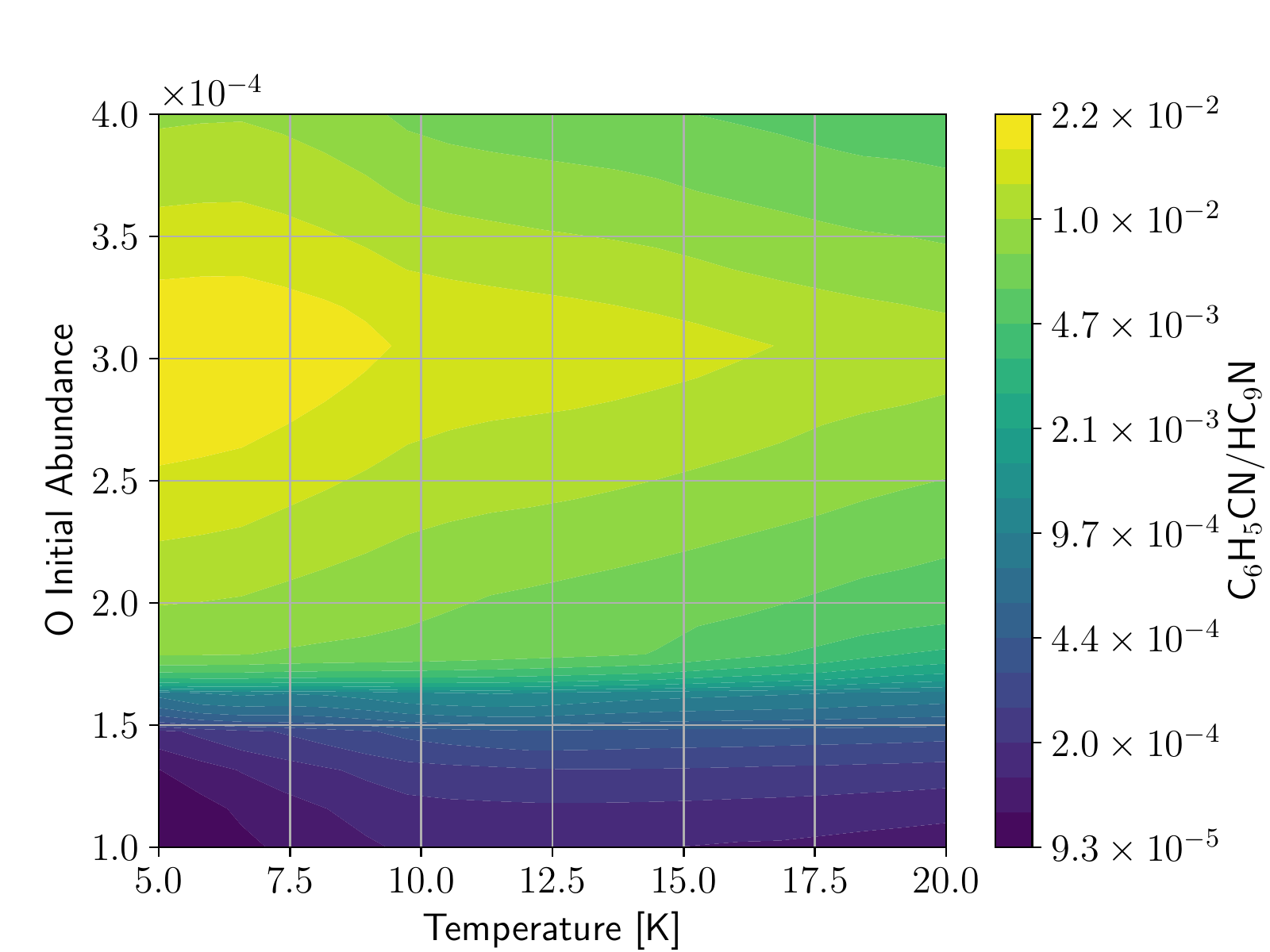}
    \includegraphics[width=0.49\textwidth]{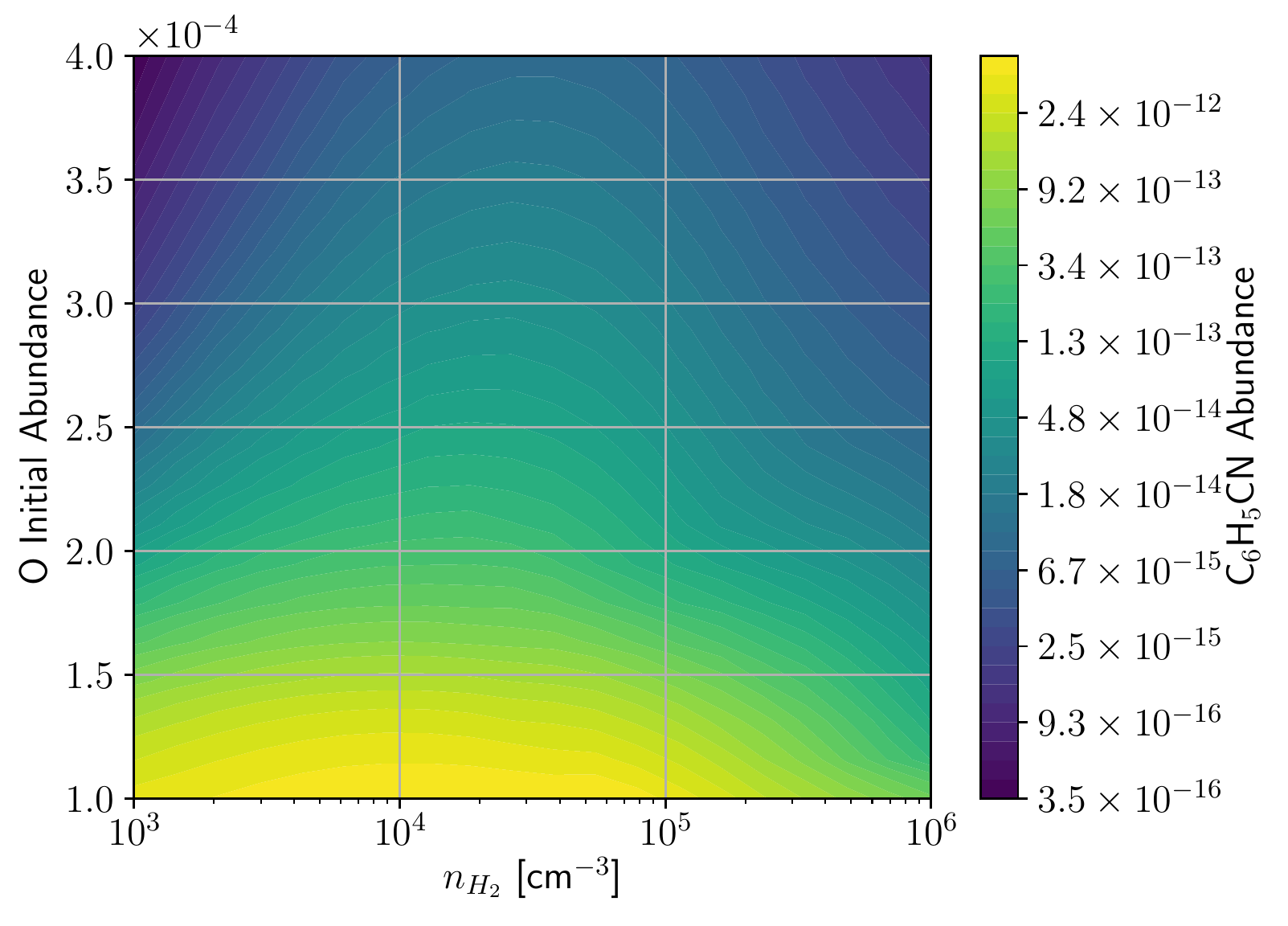}
    \includegraphics[width=0.49\textwidth]{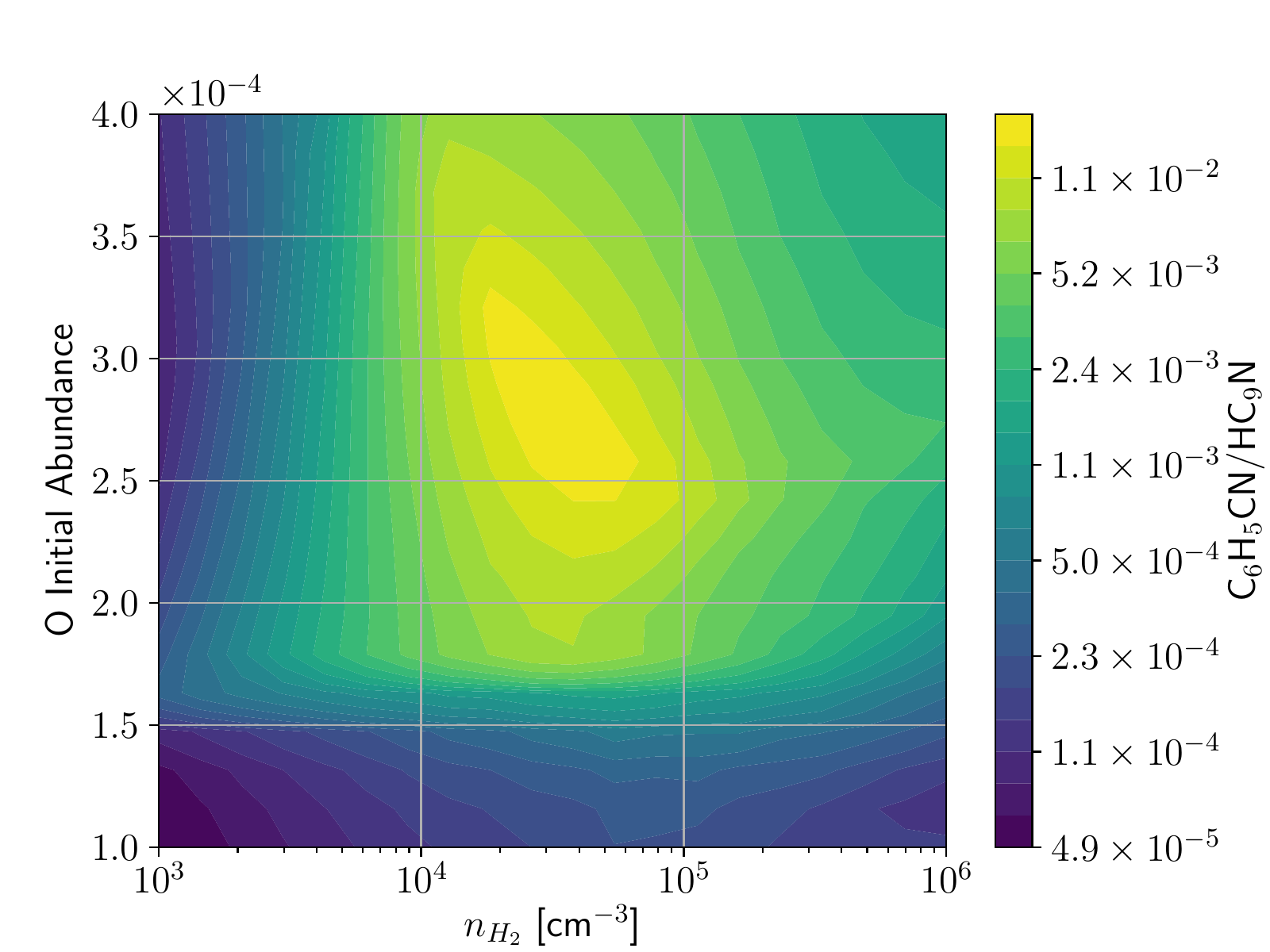}
    \caption{Simulated \ce{C6H5CN} abundance (\emph{Left}) and \ce{C6H5CN}/\ce{HC9N} abundance ratio (\emph{Right}) from \textsc{nautilus} chemical models over a range of gas and grain temperatures, gas densities, and initial elemental oxygen abundances.}
    \label{fig:grids}
\end{figure}

Similarities among related sources may indicate the importance of the parent cloud in determining the abundances of interstellar molecules, including aromatics. For example, if aromatic molecules are simply relics that have survived the diffuse cloud stage, their abundances should be largely uncorrelated to those of cyanopolyynes which are readily formed in dark clouds. If correct, large variations of the \ce{HC7N}/\ce{C6H5CN} might be expected. Alternatively, while no correlation was found between evolutionary stage and the benzonitrile abundance, if aromatics are produced in dark clouds but with different efficiencies compared to carbon chains, the \ce{HC9N}/\ce{C6H5CN} ratio may be considerably more sensitive to the physical conditions and initial elemental reservoirs of the clouds themselves, particularly the C/O ratio, as seen in Fig.~\ref{fig:grids}. Regardless, the presence of \ce{C6H5CN} toward the VeLLO MC27/L1521F, which displays evidence for an incipient protostar, at any significant level implies that aromatic species survive at least the initial phase of star formation.

It should be emphasized that there is considerable evidence casting doubt as to the importance of inheritance by simple aromatics, but the same may not be true for larger species. Several studies conclude that small PAHs ($<$20-30 atoms) are readily dissociated by starlight in high yield when transiting the diffuse gas \cite{chabot_coulomb_2019,rapacioli_formation_2006,montillaud_evolution_2013} because they are unable to efficiently radiatively relax following absorption of a UV photon. If \ce{C6H6} is present at the beginning of the dark cloud phase, for example, our models (Fig.~\ref{fig:obs_and_model}) require an abundance of $\sim$10$^{-5}$, or nearly 60\% of the model's reactive carbon budget (accounting for the six carbons in benzene), to reproduce the benzonitrile abundance on the timescale of prestellar sources. A similar large initial seed of cyclic species is also required to reproduce the abundance of the cyanonaphthalenes \cite{McGuire:2020aa} and cyano-cyclopentadiene \cite{McCarthy:2020aa}. Placing such a large fraction of all carbon in a single molecule appears highly unrealistic given that the estimated fraction of carbon locked up in PAHs overall is estimated to be 10-25\% \cite{Tielens:2008fx}, and because one might expect the initial aromatic reservoir to be dominated by larger, more stable cyclic species. However, the dissociation of large cyclic species in the diffuse clouds may produce modest-sized molecular fragments that conceivably could be important precursors for small aromatic species, but are not considered here. Previous studies suggest PAHs undergoing Coulomb explosion primarily produce hydrogen, acetylene (\ce{C2H2}), and carbon chains containing 10-15($\pm$3) carbon atoms, depending the internal energy and the size of the PAH \cite{montillaud_evolution_2013,chabot_coulomb_2019}. More generally, our modeling highlights the existence of a previously unknown and non-negligible reservoir of complex carbon, and one which should be considered in chemical models of protostellar evolution.

In summary, the detection of benzonitrile towards the starless cloud core TMC-1 expanded our view of aromatic chemistry beyond the initial factories of carbon-rich stars. The ease with which the same aromatic has been subsequently detected in other molecular clouds points to the generality of this chemistry in early star formation, rather than an anomaly specific to the rich carbon-chain chemistry of TMC-1. The high abundances of this aromatic molecule are far greater than what present chemical models predict, and exhibit sizable variations between clouds of roughly similar ages. For these reasons, the present observations should serve as a strong impetus for new laboratory and theoretical studies to investigate formation pathways of small aromatics in greater detail, and motivate additional observations towards objects over a variety of initial conditions or are further along the pathway to star formation.


\newpage
\section*{Methods}

\subsection*{Spectral Stacking Routine} \label{sec:stacking}

Full details of the methodology used to detect and quantify molecules in our spectra are provided in Loomis et al. \cite{Loomis:2020aa}. Briefly, we first perform a standard steepest descent fit to the molecules in our data, using model spectra generated using the formalisms outlined in Turner \cite{Turner:1991um}, which includes corrections for optical depth, and adjusted for the effects of beam dilution. The specific transition parameters for each species are obtained from spectral line catalogs primarily pulled from publicly accessible databases (\href{https://spec.jpl.nasa.gov}{\texttt{https://spec.jpl.nasa.gov}} and \href{https://cdms.astro.uni-koeln.de/}{\texttt{https://cdms.astro.uni-koeln.de/}}) or generated from spectroscopic parameters provided in their respective publications. Full details for each catalog are provided in Harvard Dataverse repositories described below. The substantial number of transitions used in the analysis makes in impractical to provide a table of parameters in-text. Instead, the interested reader is referred to the catalog files in the online supplementary data which contain all of the required information in a machine-readable format. For each source, we fit either one or two distinct velocity components ($v_{lsr}$), and simultaneously fit for column densities ($N_T$), source sizes ($\theta_s$), excitation temperatures ($T_{ex}$), and linewidth ($\Delta V$).

In the case of benzonitrile, there are many transitions covered by our spectra that are not visible above the local RMS noise level of the observations. We therefore extract a small portion of the observations centered around each spectral line. An SNR weighted average of these spectra is then performed based on the expected intensity of the line (derived from the MCMC parameters) and the local RMS noise of the observations. For the purposes of this analysis, largely due to hyperfine splitting, we treat the signals on a per-line basis rather than a per-transition basis. The result is that the stacked feature is somewhat broadened, as the hyperfine components and velocity components are not collapsed, but there is no over-counting of flux. This results in a substantial increase to overall SNR, with the spectrum now encapsulating the total information content of all observed lines, rather than only that from the brightest lines. Finally, the model spectra are stacked using identical weights, and that stacked model is used as a matched filter which is cross-correlated with the stacked observations. The resulting impulse response spectrum provides a lower limit statistical significance to the detection. Because the filter contains the same broadened hyperfine and velocity structure as the stack, there is no loss in significance. 

\sitablename~\ref{tab:spectra_table} shows the total number of transitions (including hyperfine components) of the molecules analyzed in this paper that were covered by GOTHAM and ARKHAM observations at the time of analysis and were above our predicted flux threshold of 5\%, as discussed in \cite{Loomis:2020aa}. Also included are the number of transitions, if any, that were coincident with interfering transitions of other species, and the total number of lines used after excluding interlopers. Observational data windowed around these transitions, spectroscopic properties of each transition, and the partition functions used in the MCMC analysis are provided in Harvard Dataverse repositories \cite{GOTHAMDR1, ARKHAMDR1}.

\subsection*{Astrochemical Modeling}
\label{sec:model}

A modified gas-grain kinetic chemical model, \textsc{nautilus} \cite{Ruaud:2016bv}, with adaptations beyond Ref.~\cite{McGuire:2018it} and \cite{shingledecker_cosmic-ray-driven_2018} was used to model the formation of aromatic species. We note that the model described here uses an identical chemical network and initial parameter set (see \sitablenames~\ref{tab:model_parameters} and \ref{tab:model_elements}) as the GOTHAM-related works in Refs~\cite{McGuire:2020aa,McGuire:2020bb,Xue:2020aa,McCarthy:2020aa,Loomis:2020aa} unless specifically stated otherwise. 

For this work, we have added the reactions summarized in \sitablenames~\ref{tab:newgasnetwork} and \ref{tab:newsolidnetwork}, which expand upon those presented in Ref.~\cite{McGuire:2018it}. In our models, the dominant formation pathway to benzene is through the following dissociative recombination process proposed by McEwan et al. \cite{McEwan:1999ia}: 

\begin{equation}
    \ce{C6H7+} + \ce{e-} \longrightarrow \ce{H} + \ce{C6H6}
\end{equation}

\noindent
The fact that our model underproduces benzonitrile is indicative of some deficiency in the included aromatic chemistry, such as missing formation routes or incorrect reaction rates. For example, the cyclization of linear carbon-chain species, either in the gas or on grains, represents an attractive possible formation pathway that is not currently included in our network. Moreover, only $\sim$1\% of \ce{C6H5CN} is frozen on grains in our model; however, this value could plausibly be much greater in astrophysical environments, especially if there exists a viable grain-surface formation pathway. If so, additional low-temperature desorption mechanisms, such as sputtering by cosmic ray bombardment or accelerated gas-phase particles \cite{shingledecker_cosmic-ray-driven_2018,Burkhardt:2019gf} will need to be considered in dark cloud chemistry models. Another possibility is that some of the as-yet-undetected species that have been proposed \cite{Jones:2011yc} to be important precursors to benzene - such as \ce{C6H7+}, 1,3-butadiene, and propene - may be likewise underpredicted in our model, thereby resulting in low calculated abundances of \ce{C6H6}. Interstellar observations of these potential precursors would significantly help to constrain the proposed pathways leading to benzene. 

Destruction routes that lead to other aromatic or cyclic molecules, such as cyanonaphthalene and cyano-cyclopentadiene have also been included in our latest model. However, expanding the network to include additional reactions with potential precursors to benzene, such as \ce{C6H7+}, 1,3-butadiene (\ce{CH2CHCHCH2}), and the propargyl (\ce{CH3CHCH2}) and phenyl (\ce{C6H5}) radicals \cite{Jones:2011yc}, did little to increase the abundance of benzonitrile.

While our models do not reproduce the derived abundances of benzonitrile, we have explored their sensitivity to certain key parameters in our simulations to study the importance of the parent cloud to the production of benzonitrile and, by proxy, other aromatics. In Fig.~\ref{fig:grids}, the calculated abundances of \ce{C6H5CN} and the abundance ratios of \ce{C6H5CN}/\ce{HC9N} are shown over a range of temperatures, densities, and initial oxygen abundance. From this plot, there exists an optimal temperature (10\,K) and density ($\sim$10$^4$~cm$^{-3}$) that maximizes the \ce{C6H5CN} abundance, which is in agreement with our model's initial parameters (\sitablename~\ref{tab:model_parameters}). Thus, the observed aromatic chemistry is much richer than what our models predict over a wide range of physical conditions (Fig.~\ref{fig:grids} \emph{Left}). As expected, the formation of both long carbon-chains and cyclic species is strongly correlated with the C/O ratio, as an increase in available C enhances the abundance of many carbon-bearing species. When the initial oxygen abundance with respect to hydrogen is $\sim$2-3.5$\times$10$^{-4}$, resulting in a carbon-poor ratio (C/O $\sim$0.5-0.85), the production of \ce{C6H5CN} relative to \ce{HC9N} becomes more efficient. This is perhaps because the C/O ratio becomes increasingly more important as one considers the formation of longer carbon chains. Meanwhile, the dominant benzene pathway in this model does not require the presence of chains larger than five carbons. Similar to the C/O ratio, relative \ce{C6H5CN} production appears to be favored compared to \ce{HC9N} in denser ($\sim$5$\times$10$^5$\,cm$^{-3}$) gas, likely due to higher collision rates increasing the efficiency of the ion-neutral reactions in the benzene pathway and the destruction of the longest, and thus least stable, carbon chains. These scenarios underscore that it is feasible to produce significant variations between the abundance of cyanopolyynes and cyclic species simply by changing the initial conditions, which could explain the different \ce{HC9N}/\ce{C6H5CN} ratios seen toward Taurus and Serpens.

\clearpage

\bibliography{bibliography,shingledecker,burkhardt}
\bibliographystyle{naturemag}


\section*{Contributions}
All authors contributed to the design of the GOTHAM and ARKHAM survey, and helped revise the manuscript.
A.M.B and B.A.M. performed the astronomical observations and subsequent analysis. 
R.A.L., K.L.K.L., and B.A.M. performed the spectral fitting analyses.
A.M.B. and C.N.S. contributed or undertook the astronomical modeling and simulations. 
A.M.B., M.C.M. and B.A.M. wrote the manuscript with the help of C.N.S.

\section*{Data statement}
The datasets analyzed during the current study are available in the Green Bank Telescope archive (\href{https://archive.nrao.edu/archive/advquery.jsp}{\texttt{https://archive.nrao.edu/archive/advquery.jsp}}).  A user manual for their reduction and analysis is available as well (\href{https://greenbankobservatory.org/science/gbt-observers/visitor-facilities-policies/data-reduction-gbt-using-idl/}{https://greenbankobservatory.org/science/gbt-observers/visitor-facilities-policies/data-reduction-gbt-using-idl/}).  For the ARKHAM survey, the complete, reduced survey data is available in the Harvard Dataverse Archive \cite{ARKHAMDR1}. For the GOTHAM survey, the complete, reduced survey data at X-band is available as Supplementary Information in \cite{McGuire:2020bb}. The individual portions of reduced spectra used in the analysis of the individual species presented here is available in the Harvard Dataverse Archive \cite{GOTHAMDR1}.
\newline

\section*{Code statement}
All the codes used in the MCMC fitting and stacking analysis presented in this paper are open source and publicly available at \href{https://github.com/ryanaloomis/TMC1\_mcmc\_fitting}{https://github.com/ryanaloomis/TMC1\_mcmc\_fitting}.
\newline

\section*{Competing Interests Statement}
The authors declare no competing interests.

\section*{Acknowledgements}
A.M.B. acknowledges support from the Smithsonian Institution as a Submillimeter Array (SMA) Fellow. C.N.S. thanks the Alexander von Humboldt Stiftung/Foundation for their generous support. A.M.B. and C.N.S. would like to also thank V. Wakelam for use of the \textsc{nautilus} v1.1 code. M.C.M and K.L.K.L. acknowledge financial support from NSF grants AST-1908576, AST-1615847, and NASA grant 80NSSC18K0396. Support for B.A.M. was provided by NASA through Hubble Fellowship grant \#HST-HF2-51396 awarded by the Space Telescope Science Institute, which is operated by the Association of Universities for Research in Astronomy, Inc., for NASA, under contract NAS5-26555. The National Radio Astronomy Observatory is a facility of the National Science Foundation operated under cooperative agreement by Associated Universities, Inc. The Green Bank Observatory is a facility of the National Science Foundation operated under cooperative agreement by Associated Universities, Inc. We wish to additionally thank the anonymous reviewers whose comments helped improve the manuscript.

\section*{Corresponding author and requests for materials}

Correspondence to A.M. Burkhardt and B.A. McGuire

 
\clearpage
 
\part*{Supplementary Information}

\renewcommand{\thefigure}{\arabic{figure}}
\renewcommand{\thetable}{\arabic{table}}
\renewcommand{\theequation}{\arabic{equation}}
\renewcommand{\figurename}{Supplementary Figure}
\renewcommand{\tablename}{Supplementary Table}
\setcounter{figure}{0}
\setcounter{table}{0}
\setcounter{equation}{0}

\section*{Observations}
\begin{figure}[!t]
    \centering
    \includegraphics[width=\textwidth]{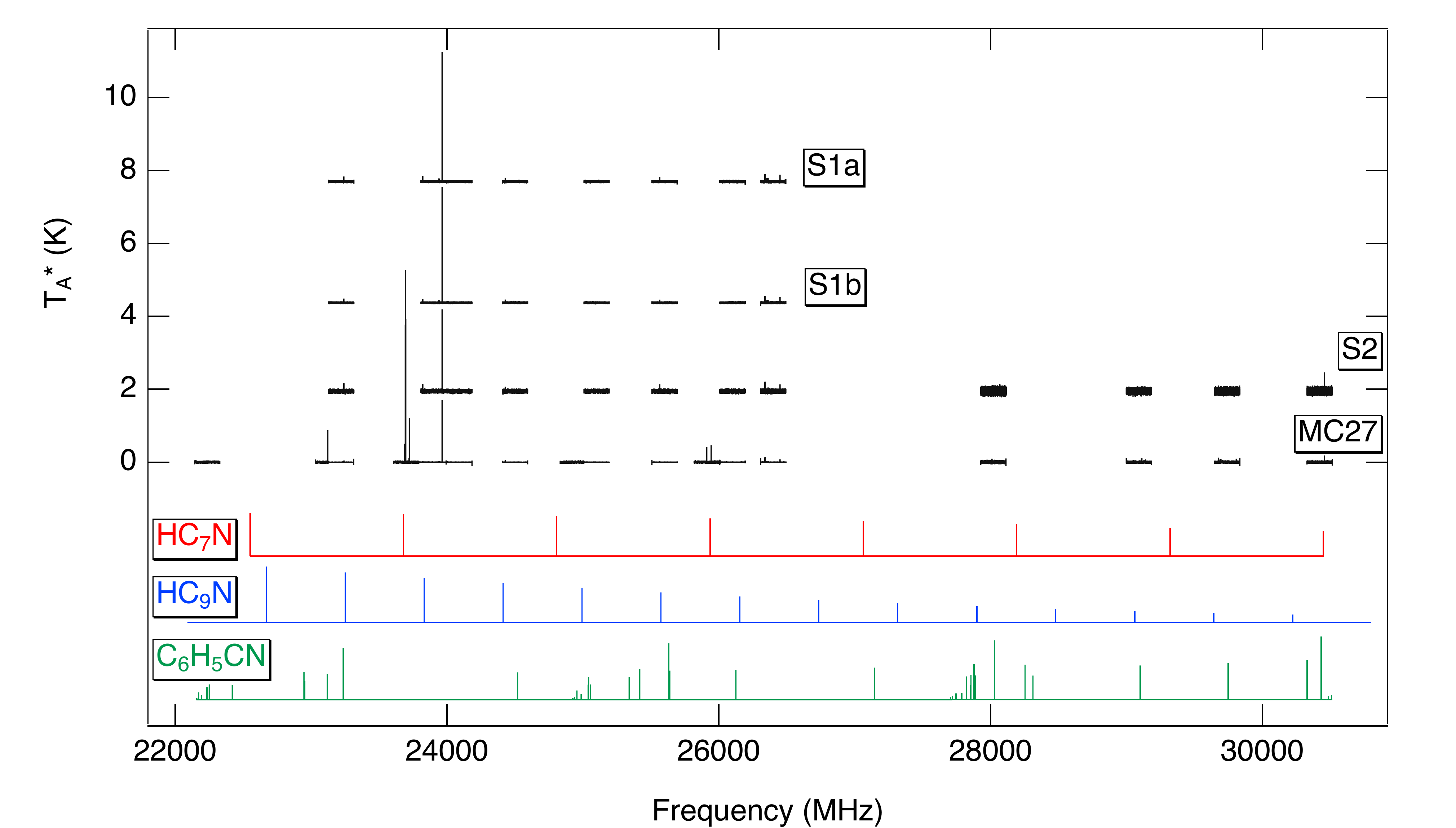}
    \caption{Full ARKHAM spectra for S1a, S1b, S2, and MC27/L1521F with example spectra of \ce{HC7N}, \ce{HC9N}, and \ce{C6H5CN} across the relevant coverage, assuming a source size of 50$^{\prime\prime}$, $T_{ex}$~=~6~K, $v_{lsr}$~=~0~km~s$^{-1}$, and $\Delta V$~=~0.15~km~s$^{-1}$. For ease of visualization, the spectra of the different ARKHAM sources are offset vertically and the simulated spectra were scaled by an arbitrary column density. For the full spectra of TMC-1, see the summary of the GOTHAM survey in Ref~\cite{McGuire:2020bb}.}
    \label{fig:full_spectra}
\end{figure}

Utilizing the seven-pixel focal plane array KFPA-band receiver, the eight IFs of the VEGAS spectrometer backend \cite{Roshi:2012he} were tuned between 23 and 26.5 GHz with 1.4 kHz spectral resolution across 187.5 MHz per IF for S1a, S1b, S2. MC27/L1521F also received additional observations at additional turnings between 22-26 GHz. The ARKHAM observations were observed in January 2018 (GBT17B-447), October 2018 - January 2019 (GBT18B-004), and January - March 2019 (GBT19A-009). The Ka-band dual-beam receiver was also utilized between December 2019 - January 2020 (GBT19B-116), with four VEGAS IFs tuned between 28 and 30.5 GHz with 1.4 kHz spectral resolution across 187.5 MHz bandwidth per IF for S2 and MC27/L1521F. The observational strategy is similar to those in the GOTHAM project \cite{McGuire:2020bb}, but with different sources (\sitablename~\ref{tab:observations_table}) and spectral tunings (\sitablename~\ref{tab:spectra_table} and \sifigurename~\ref{fig:full_spectra}). Observations were performed in ON-OFF position-switched mode with 2 minutes on target and 2 minutes at an off position throw of 1$^{\circ}$ away. The primary beam of the GBT at 23~GHz ranged between $\sim$31.5-38$^{\prime\prime}$. The K-band Focal Plane Array \cite{Morgan:2008kb} and Ka-band dual-beam receiver were used with the VEGAS spectrometer backend \cite{Roshi:2012he} configured to provide 187.5 MHz total bandwidth in each of ten target windows at a 1.4 kHz (0.02 km s$^{-1}$) spectral resolution, sufficient to put at least seven points across even our narrowest lines ($\Delta V\sim$0.15 km s$^{-1}$). The resulting spectra were placed on the atmosphere-corrected antenna temperature (T$_{\text{A}}^*$) scale \cite{Ulich:1976yt}. Data reduction was performed using the GBTIDL software package (\url{http://gbtidl.nrao.edu/}) in a similar procedure as described in \cite{McGuire:2020bb}. The receivers are primarily calibrated by means of an internal noise diode, which we assume has absolute flux density calibration uncertainty of, at best, $\sim$30\%. The spectra were averaged using a weighting scheme which corrects for the measured system temperature (T$_\text{sys}$) during each 240 s ON-OFF position cycle. A polynomial fit was used to correct for baseline fluctuations. The final root-mean-square (RMS) noise varied for the ARKHAM sources from $\sim$5-40 mK across the observations. The highest RMS values were generally at the higher frequency observations, particularly the Ka-band receiver. This prevented the individual detection of benzonitrile transitions in across these spectral ranges. For TMC-1, the RMS noise varied from $\sim$2--20~mK across the observations, as described in greater detail in \cite{McGuire:2020bb}.

\begin{table*}[!hb]
\centering
\caption{Overview of ARKHAM Targets}
\begin{tabular}{c c c c c c}
\toprule
Source& $\alpha$(J2000) & $\delta$(J2000)& $N_{\ce{H2}}$ & Evolutionary 		\\
    &	[hh:mm:ss.s]&[dd:mm:ss.s]	 & [cm$^{-2}$]	& Stage \\
\midrule
\hspace{0.1em}\vspace{-0.5em}\\
S2	&   18:30:43.8	&   -2:05:51.0	  & 1.48$\times$10$^{22\dagger}$ & Intermediate Dark Cloud\\
S1a	&   18:29:57.9  &	-1:56:19.0   & 2.78$\times$10$^{22\dagger}$ & Evidence of Collapse\\
S1b	&   18:29:57.0	&	-1:58:55.0	 & 7.33$\times$10$^{22\dagger}$ & Evidence of Collapse\\
TMC-1 & 04:41:42.5  &   +25:41:27.0   & 1.00$\times$10$^{22\ddagger}$ & Dense Cold Core  \\ 
MC27/L1521F	& 04:28:39.3  &	+26:51:39.0	 & 1.68$\times$10$^{22\oplus}$ & VeLLO\\
\bottomrule
\end{tabular}

\begin{minipage}{0.75\textwidth}
	\footnotesize
	\textbf{Note} -- 
	$^\dagger$ Friesen et al., \cite{Friesen:2013ii}\\
	$^\ddagger$  Cernicharo et al. \cite{Cernicharo:2018bv}\\
	$^{\oplus}$ Chitsazzadeh \cite{Chitsazzadeh2014PhDT}
\end{minipage}

\label{tab:observations_table}

\end{table*}

\begin{table*}
\centering
\caption{Overview of ARKHAM Spectral Setups}
\begin{tabular}{c c c c c c}
\toprule
Source & Frequency Coverage & Average RMS & \ce{C6H5CN} Lines & Excluded   & Lines Used \\
           & [GHz]          & [mK] & $>$Threshold$^\dagger$ & Lines$^\ddagger$ &  in MCMC$^\ddagger$ \\
\midrule
\hspace{0.1em}\vspace{-0.5em}\\
S2	          &23.13-23.31, 23.81-24.18,  & 24.8 & 30 & 0 & 30\\
&24.41-24.59, 25.01-25.19, \\
&25.51-25.69, 26.31-26.49,\\
&27.93-28.11, 29.00-29.18, \\
&29.65-29.83, 30.33-30.51\\
\hline
S1a	          &23.13-23.31, 23.81-24.18, & 11.3 & 21 & 0 & 21 \\
 & 24.41-24.59, 25.01-25.19, \\
 & 25.51-25.69, 26.31-26.49\\
 \hline
S1b	           & 23.13-23.31, 23.81-24.18, & 10.2 & 21 & 0 & 21\\
 & 24.41-24.59, 25.01-25.19,\\
 & 25.51-25.69, 26.31-26.49\\
 \hline
TMC-1      & $\oplus$ & 2-20$^{\oplus}$ & 255$^{\odot}$ & 0$^{\odot}$ & 255$^{\odot}$\\ 
\hline
MC27/  & 22.14-22.33, 23.03-23.31, & 9.1 & 66 & 0 & 66\\
L1521F & 23.61-24.18, 24.41-24.59, \\
& 24.83-25.19, 25.51-25.69,\\
& 25.82-26.00, 26.31-26.49, \\
& 27.93-28.11, 29.00-29.18, \\
& 29.65-29.83, 30.33-30.51\\
\bottomrule
\end{tabular}

\begin{minipage}{0.75\textwidth}
	\footnotesize
	\textbf{Note} -- Frequency coverage are rounded to nearest 10 MHz. The full line list catalog for \ce{HC7N}, \ce{HC9N}, and \ce{C6H5CN} and full spectrum for S1a, S1b, S2, and MC27/L1521F can be found at \cite{ARKHAMDR1}.\\
	$\oplus$ For full discussion of spectral coverage and RMS of TMC-1, see the summary of the GOTHAM survey in Ref~\cite{McGuire:2020bb}. \\
	$\odot$ For full discussion for the MCMC fitting of benzonitrile toward TMC-1, see the summary in Ref~\cite{Loomis:2020aa}.\\ 
	$\dagger$ A 5\% peak threshold was applied among all \ce{C6H5CN} transitions within the observed spectral coverage to filter which would be used in the MCMC fit (see the Spectral Stacking Routine section of Methods and Ref.~\cite{Loomis:2020aa}) \\
	$\ddagger$ Transitions with any interloping features were excluded from the MCMC analysis (see the Spectral Stacking Routine section of Methods and Ref.~\cite{Loomis:2020aa})
\end{minipage}

\label{tab:spectra_table}

\end{table*}

\clearpage
\newpage{}

\begin{table*}
\centering
\caption{Individually Detected Benzonitrile Transitions}
\begin{tabular}{c c c c c}
\toprule
\multicolumn{2}{c}{Transition}	& Frequency & $E_U$& $S_{ij}\mu^2$	\\
$J^{\prime}_{K_a,K_c} - J^{\prime\prime}_{K_a,K_c}$	& $F^{\prime}-F^{\prime\prime}$      &	[MHz]     &[K]	 & [Debye$^2$]	 \\
\midrule
\hspace{0.1em}\vspace{-0.5em}\\
9$_{0,8}-8_{0,8}$   & $8-7$     & 23227.6903(3) & 5.72  & 33.1\\
                    & $10-9$    & 23227.7105(3) & 5.72  & 41.5\\
                    & $8-8$     & 23227.7127(3) & 5.72  & 37.1\\
10$_{0,10}-9_{0,9}$ & $9-8$     & 25622.3093(4) & 6.95  & 37.3\\
                    & $11-10$   & 25622.3253(4) & 6.95  & 45.7\\
                    & $10-9$    & 25622.3260(4) & 6.95  & 41.3\\
9$_{1,8}-8_{1,7}$   & $8-7$     & 25630.0360(4) & 6.44  & 19.7\\
                    & $9-8$     & 25630.0498(4) & 6.44  & 22.0\\
                    & $10-9$    & 25630.0560(4) & 6.44  & 24.7\\
\bottomrule
\end{tabular}

\begin{minipage}{0.75\textwidth}
	\footnotesize
	\textbf{Note} -- Statistical uncertainties (1$\sigma$), derived from the best-fitting constants are given in parentheses in units of the last significant digit \cite{McGuire:2018it}. The full catalog of lines used in stacking and MCMC analysis for ARKHAM and GOTHAM are provided in Ref.~\cite{ARKHAMDR1} and \cite{GOTHAMDR1}, respectively.
\end{minipage}
\label{tab:bn_transitions}

\end{table*}

\clearpage
\newpage{}
\section*{Benzonitrile Analysis Results}

\subsection*{Serpens 2}
\begin{figure}
    \centering
    \includegraphics[width=\textwidth]{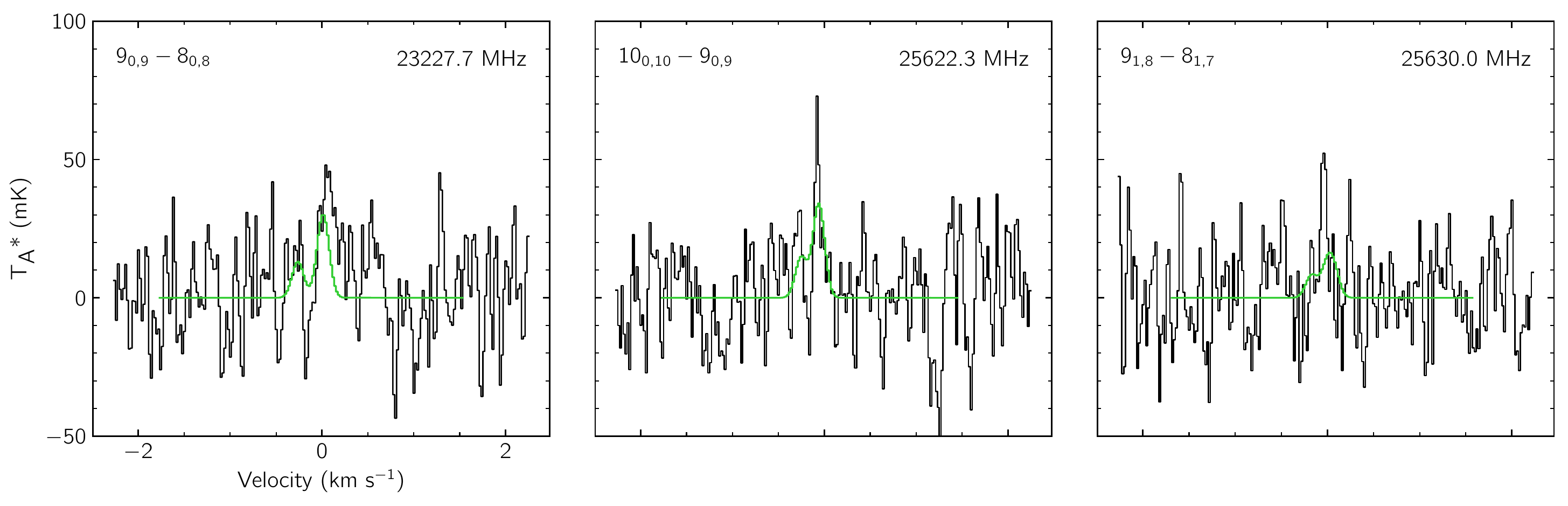}
    \caption{Individual line detections of benzonitrile in the Serpens 2 data. The spectra (black) are displayed in velocity space relative to 7.117\,km\,s$^{-1}$, and using the rest frequency given in the top right of each panel. Quantum numbers are given in the top left of each panel, neglecting hyperfine splitting. The best-fit model to the data is overlaid in green. See \sitablename~\ref{s2_bn_results}.}
    \label{s2_bn_spectra}
\end{figure}

\begin{figure}
\centering
\includegraphics[width=\textwidth]{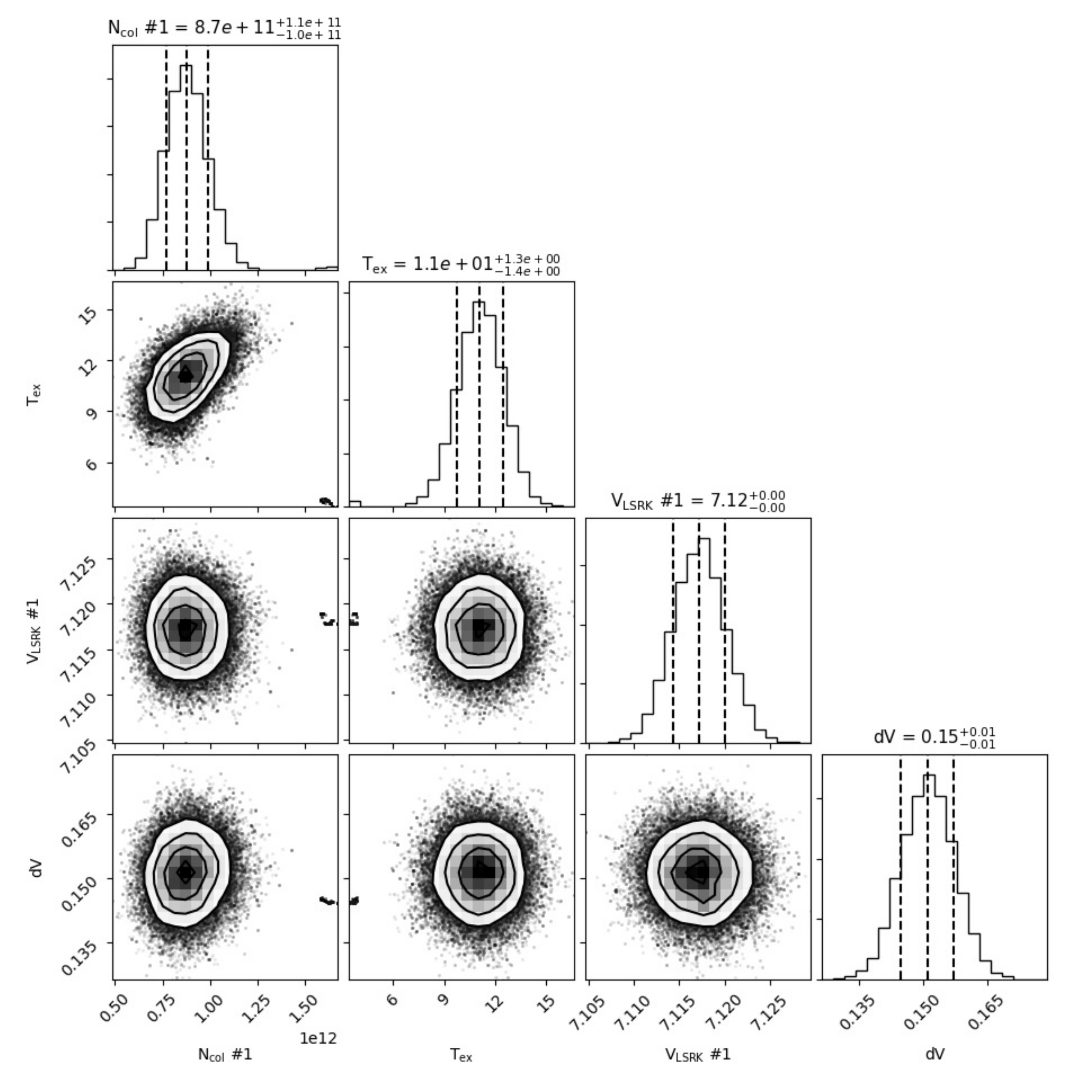}
\caption{Parameter covariances and marginalized posterior distributions for the benzonitrile MCMC fit for Serpens 2. 16$^{th}$, 50$^{th}$, and 84$^{th}$ confidence intervals (corresponding to $\pm$1 sigma for a Gaussian posterior distribution) are shown as vertical lines. }
\label{s2_bn_triangle}
\end{figure}

The resulting best-fit parameters for the MCMC analysis of benzonitrile in Serpens 2 are given in \sitablename~\ref{s2_bn_results}. Detected individual lines are shown in \sifigurename~\ref{s2_bn_spectra}, while the stacked spectrum and matched filter results are shown in Figure~\ref{fig:bn_stack}c. A corner plot of the parameter covariances for the benzonitrile MCMC fit is shown in \sifigurename~\ref{s2_bn_triangle}. The largest source of error was a result of the degeneracy between the column density and the excitation temperature ($\sim$12\%), compared to the much lower uncertainties of $v_{lsr}$ ($\sim$0.04\%) and $\Delta V$ ($\sim$4\%) facilitated by the high spectral resolution and narrow line widths of our observations.

\begin{table}[!hb]
\centering
\caption{Benzonitrile best-fit parameters for Serpens 2 from MCMC analysis}
\begin{tabular}{c c c c c c}
\toprule
\multirow{2}{*}{Component}&	$v_{lsr}$					&	Size					&	\multicolumn{1}{c}{$N_T^\dagger$}					&	$T_{ex}$							&	$\Delta V$		\\
			&	(km s$^{-1}$)				&	($^{\prime\prime}$)		&	\multicolumn{1}{c}{(10$^{11}$ cm$^{-2}$)}		&	(K)								&	(km s$^{-1}$)	\\
\midrule
\hspace{0.1em}\vspace{-0.5em}\\
C1	&	$7.117^{+0.003}_{-0.003}$	&	[$400$]	&	$8.75^{+1.12}_{-1.04}$	&	$11.1^{+1.3}_{-1.4}$	&	$0.151^{+0.006}_{-0.006}$\\
\hspace{0.1em}\vspace{-0.5em}\\
\midrule
$N_T$ (Total)$^{\dagger\dagger}$	&	 \multicolumn{5}{c}{$8.75^{+1.12}_{-1.04}\times 10^{11
}$~cm$^{-2}$}\\
\bottomrule
\end{tabular}

\begin{minipage}{0.75\textwidth}
	\footnotesize
	\textbf{Note} -- The quoted uncertainties represent the 16$^{th}$ and 84$^{th}$ percentile ($1\sigma$ for a Gaussian distribution) uncertainties. \\
	$^\dagger$See \sifigurename~\ref{s2_bn_triangle} for a covariance plot, and  Loomis et al.\cite{Loomis:2020aa} for a detailed explanation of the methods used to constrain these quantities and derive the uncertainties.\\
	$^{\dagger\dagger}$Uncertainties derived by adding the uncertainties of the individual components in quadrature.
\end{minipage}

\label{s2_bn_results}

\end{table}

\clearpage
\newpage{}
\subsection*{Serpens 1a}
\begin{figure}[!t]
    \centering
    \includegraphics[height=0.25\textheight]{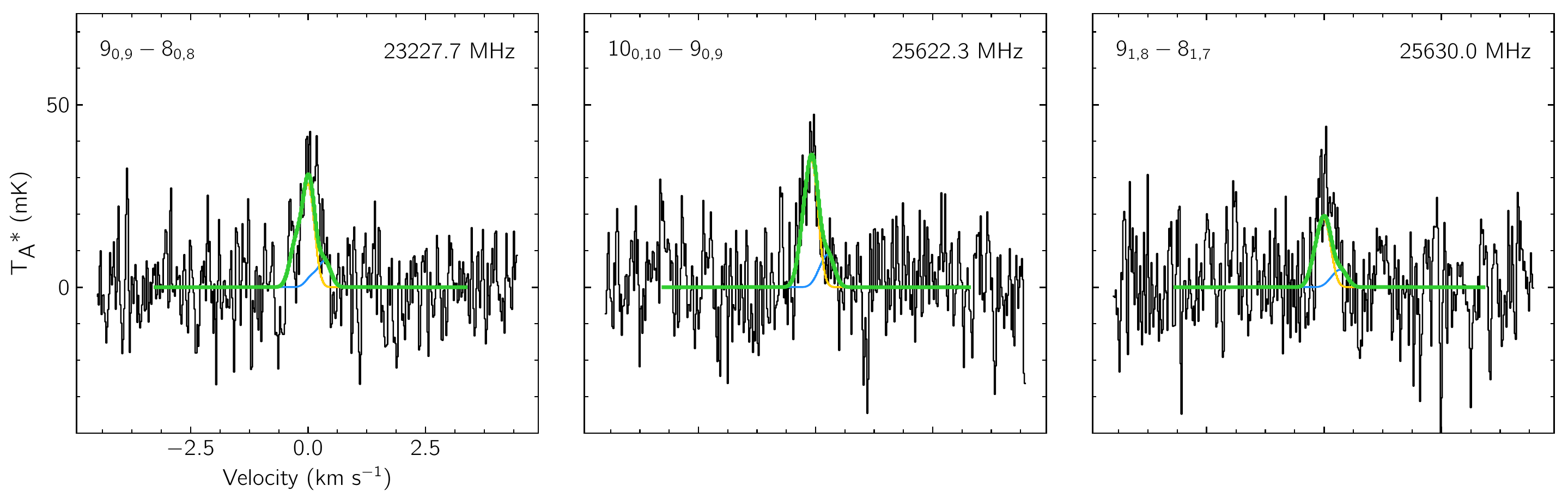}
    \caption{Individual line detections of benzonitrile in the Serpens 1a data. The spectra (black) are displayed in velocity space relative to 7.67\,km\,s$^{-1}$, and using the rest frequency given in the top right of each panel. Quantum numbers are given in the top left of each panel, neglecting hyperfine splitting. The best-fit model to the data, including all velocity components, is overlaid in green. Simulated spectra of the individual velocity components are shown in: blue (7.32\,km\,s$^{-1}$) and gold (7.67\,km\,s$^{-1}$). See \sitablename~\ref{s1a_bn_results}.}
    \label{s1a_bn_spectra}
\end{figure}

\begin{figure}[!b]
\centering
\includegraphics[width=\textwidth]{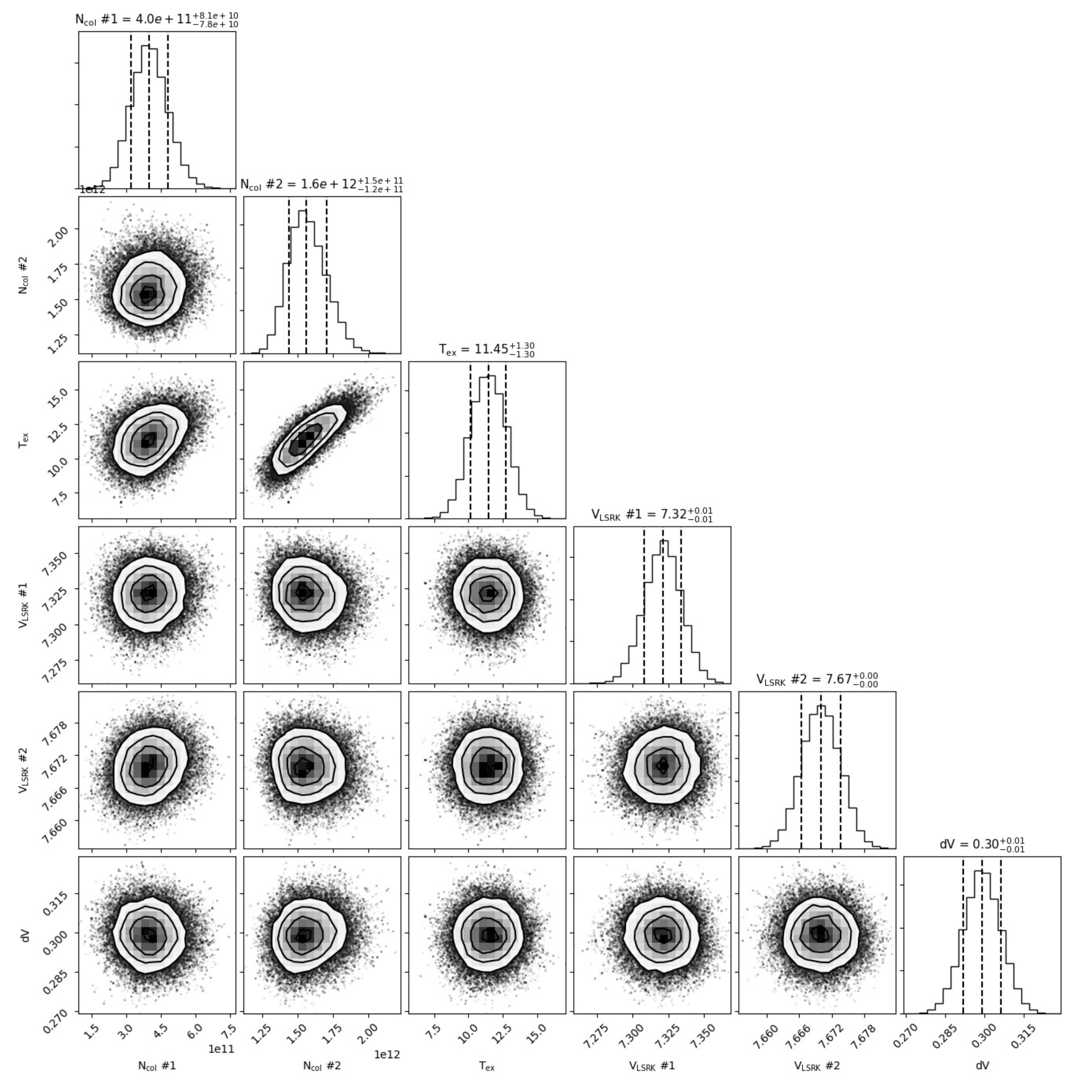}
\caption{Parameter covariances and marginalized posterior distributions for the benzonitrile MCMC fit for Serpens 1a. 16$^{th}$, 50$^{th}$, and 84$^{th}$ confidence intervals (corresponding to $\pm$1 sigma for a Gaussian posterior distribution) are shown as vertical lines. }
\label{s1a_bn_triangle}
\end{figure}

The resulting best-fit parameters for the MCMC analysis of benzonitrile in S1a are given in \sitablename~\ref{s1a_bn_results}. Detected individual lines are shown in \sifigurename~\ref{s1a_bn_spectra}, while the stacked spectrum and matched filter results are shown in Figure~\ref{fig:bn_stack}a. A corner plot of the parameter covariances for the benzonitrile MCMC fit is shown in \sifigurename~\ref{s1a_bn_triangle}. The largest source of error was a result of the degeneracy between the column density and the excitation temperature ($\sim$10-20\%), compared to the much lower uncertainties of $v_{lsr}$ ($\sim$0.05-0.2\%) and $\Delta V$ ($\sim$2\%) facilitated by the high spectral resolution and narrow line widths of our observations.

\begin{table}[htb!]
\centering
\caption{Benzonitrile best-fit parameters for Serpens 1a from MCMC analysis}
\begin{tabular}{c c c c c c}
\toprule
\multirow{2}{*}{Component}&	$v_{lsr}$					&	Size					&	\multicolumn{1}{c}{$N_T^\dagger$}					&	$T_{ex}$							&	$\Delta V$		\\
			&	(km s$^{-1}$)				&	($^{\prime\prime}$)		&	\multicolumn{1}{c}{(10$^{12}$ cm$^{-2}$)}		&	(K)								&	(km s$^{-1}$)	\\
\midrule
\hspace{0.1em}\vspace{-0.5em}\\
C1	&	$7.321^{+0.013}_{-0.013}$	&	[$400$]	&	$0.40^{+0.08}_{-0.08}$	&	\multirow{3}{*}{$11.4^{+1.3}_{-1.3}$}	&	\multirow{3}{*}{$0.299^{+0.007}_{-0.007}$}\\
\hspace{0.1em}\vspace{-0.5em}\\
C2	&	$7.670^{+0.004}_{-0.004}$	&	[$400$]	&	$1.56^{+0.15}_{-0.12}$	&		&	\\
\hspace{0.1em}\vspace{-0.5em}\\
\midrule
$N_T$ (Total)$^{\dagger\dagger}$	&	 \multicolumn{5}{c}{$1.96^{+0.17}_{-0.15}\times 10^{12}$~cm$^{-2}$}\\
\bottomrule
\end{tabular}
\begin{minipage}{0.75\textwidth}
	\footnotesize
	\textbf{Note} -- The quoted uncertainties represent the 16$^{th}$ and 84$^{th}$ percentile ($1\sigma$ for a Gaussian distribution) uncertainties. \\
	$^\dagger$See \sifigurename~\ref{s1a_bn_triangle} for a covariance plot, and  Loomis et al.\cite{Loomis:2020aa} for a detailed explanation of the methods used to constrain these quantities and derive the uncertainties.\\
	$^{\dagger\dagger}$Uncertainties derived by adding the uncertainties of the individual components in quadrature.
\end{minipage}

\label{s1a_bn_results}

\end{table}

\newpage

\newpage

\clearpage
\newpage{}
\subsection*{Serpens 1b}

\begin{figure}[!hb]
    \centering
    \includegraphics[height=0.25\textheight]{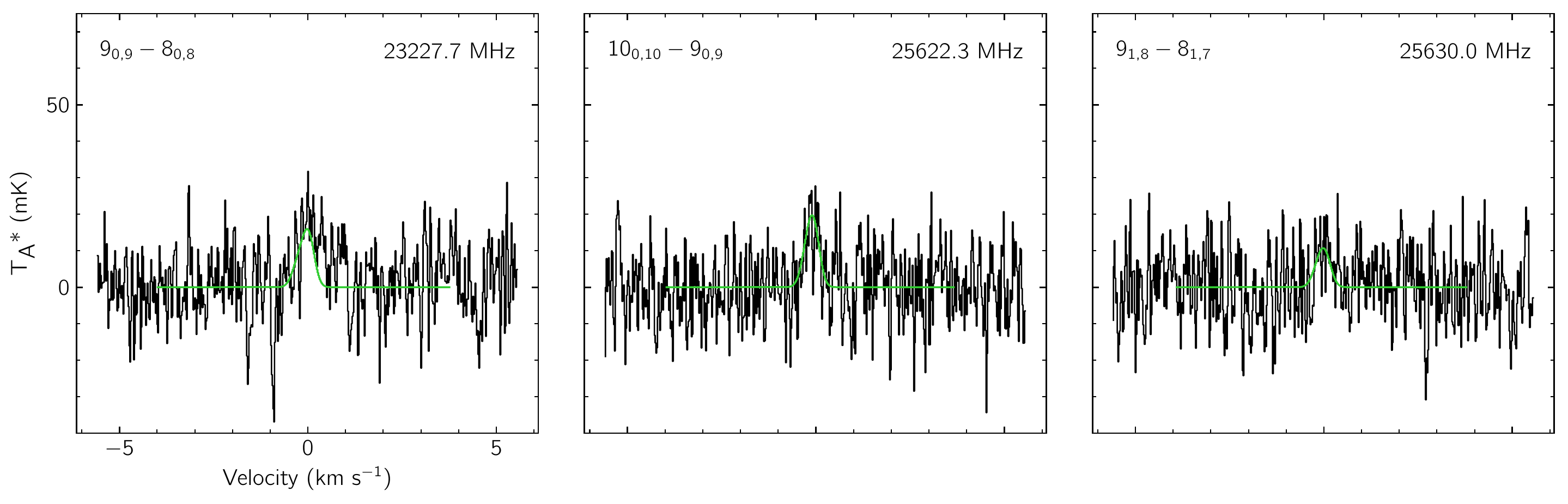}
    \caption{Individual line detections of benzonitrile in the Serpens 1b data. The spectra (black) are displayed in velocity space relative to 7.509\,km\,s$^{-1}$, and using the rest frequency given in the top right of each panel. Quantum numbers are given in the top left of each panel, neglecting hyperfine splitting. The best-fit model to the data is overlaid in green. See \sitablename~\ref{s1b_bn_results}.}
    \label{s1b_bn_spectra}
\end{figure}

\begin{figure}
\centering
\includegraphics[width=\textwidth]{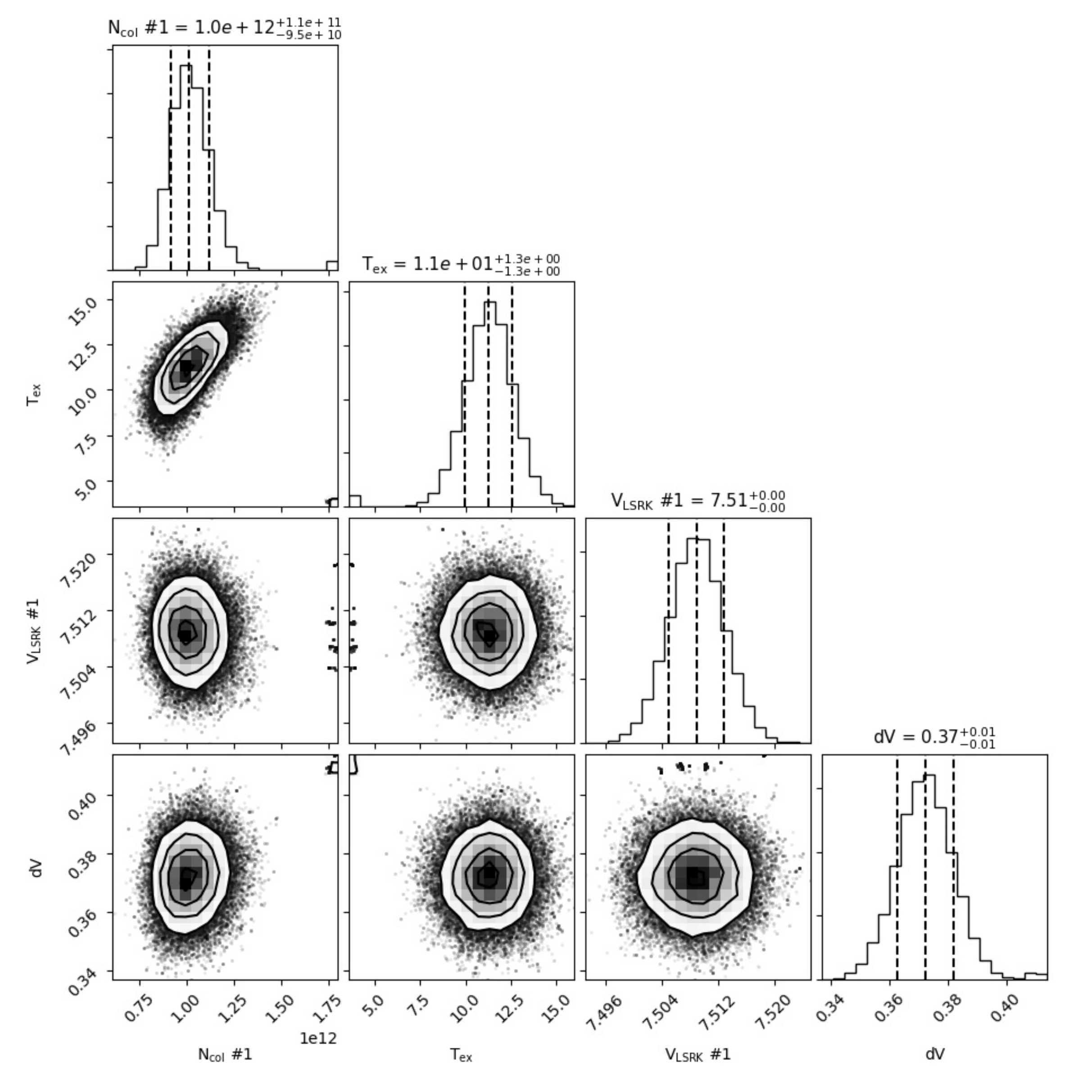}
\caption{Parameter covariances and marginalized posterior distributions for the benzonitrile MCMC fit for Serpens 1b. 16$^{th}$, 50$^{th}$, and 84$^{th}$ confidence intervals (corresponding to $\pm$1 sigma for a Gaussian posterior distribution) are shown as vertical lines. }
\label{s1b_bn_triangle}
\end{figure}

The resulting best-fit parameters for the MCMC analysis of benzonitrile in S1b are given in \sitablename~\ref{s1b_bn_results}. Detected individual lines are shown in \sifigurename~\ref{s1b_bn_spectra}, while the stacked spectrum and matched filter results are shown in Figure~\ref{fig:bn_stack}b. A corner plot of the parameter covariances for the benzonitrile MCMC fit is shown in \sifigurename~\ref{s1b_bn_triangle}.  The largest source of error was a result of the degeneracy between the column density and the excitation temperature ($\sim$10\%), compared to the much lower uncertainties of $v_{lsr}$ ($\sim$0.05\%) and $\Delta V$ ($\sim$2\%) facilitated by the high spectral resolution and narrow line widths of our observations.

\begin{table}[htb!]
\centering
\caption{Benzonitrile best-fit parameters for Serpens 1b from MCMC analysis}
\begin{tabular}{c c c c c c}
\toprule
\multirow{2}{*}{Component}&	$v_{lsr}$					&	Size					&	\multicolumn{1}{c}{$N_T^\dagger$}					&	$T_{ex}$							&	$\Delta V$		\\
			&	(km s$^{-1}$)				&	($^{\prime\prime}$)		&	\multicolumn{1}{c}{(10$^{12}$ cm$^{-2}$)}		&	(K)								&	(km s$^{-1}$)	\\
\midrule
\hspace{0.1em}\vspace{-0.5em}\\
C1	&	$7.509^{+0.004}_{-0.004}$	&	[$400$]	&	$1.02^{+0.11}_{-0.09}$	&	$11.3^{+1.3}_{-1.3}$	&	$0.372^{+0.010}_{-0.009}$\\
\hspace{0.1em}\vspace{-0.5em}\\
\midrule
$N_T$ (Total)$^{\dagger\dagger}$	&	 \multicolumn{5}{c}{$1.02^{+0.11}_{-0.09}\times 10^{12}$~cm$^{-2}$}\\
\bottomrule
\end{tabular}

\begin{minipage}{0.75\textwidth}
	\footnotesize
	\textbf{Note} -- The quoted uncertainties represent the 16$^{th}$ and 84$^{th}$ percentile ($1\sigma$ for a Gaussian distribution) uncertainties. \\
	$^\dagger$See \sifigurename~\ref{s1b_bn_triangle} for a covariance plot, and  Loomis et al.\cite{Loomis:2020aa} for a detailed explanation of the methods used to constrain these quantities and derive the uncertainties.\\
	$^{\dagger\dagger}$Uncertainties derived by adding the uncertainties of the individual components in quadrature.
\end{minipage}

\label{s1b_bn_results}

\end{table}

\clearpage
\newpage{}
\subsection*{MC27/L1521F}
\begin{figure}[!b]
    \centering
    \includegraphics[width=\textwidth]{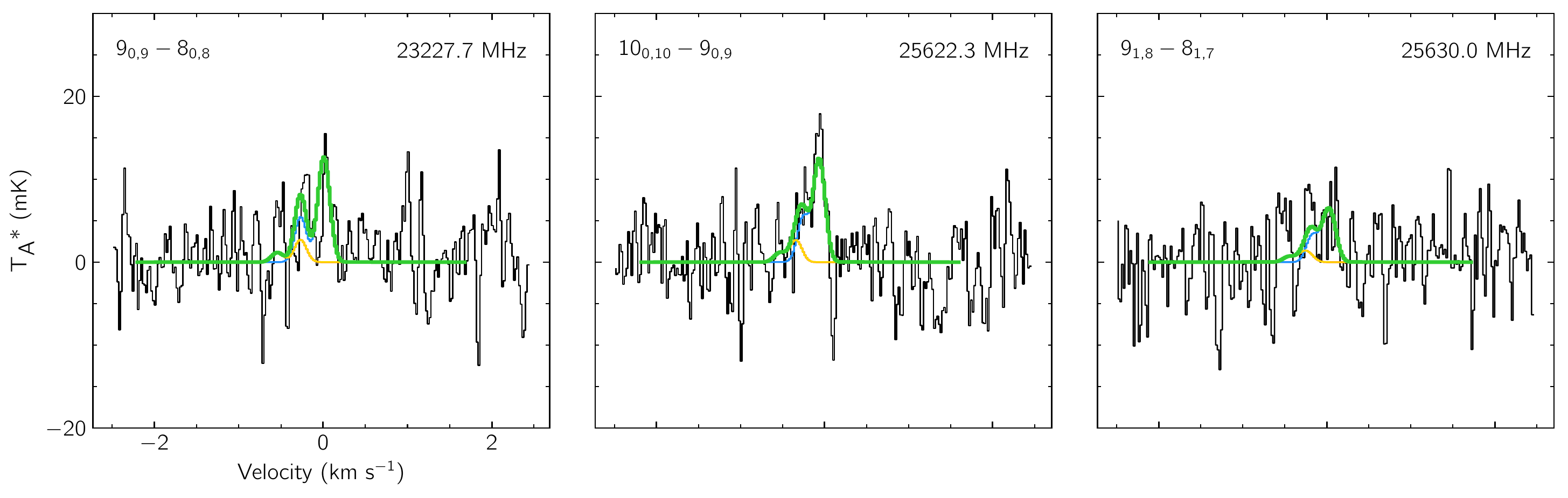}
    \caption{Individual line detections of benzonitrile in the MC27/L1521F data. The spectra (black) are displayed in velocity space relative to 6.625\,km\,s$^{-1}$, and using the rest frequency given in the top right of each panel. Quantum numbers are given in the top left of each panel, neglecting hyperfine splitting.
    The best-fit model to the data, including all velocity components, is overlaid in green. Simulated spectra of the individual velocity components are shown in: blue (6.341\,km\,s$^{-1}$) and gold (6.625\,km\,s$^{-1}$). See \sitablename~\ref{mc27_bn_results}.}
    \label{mc27_bn_spectra}
\end{figure}

\begin{figure}
\centering
\includegraphics[width=\textwidth]{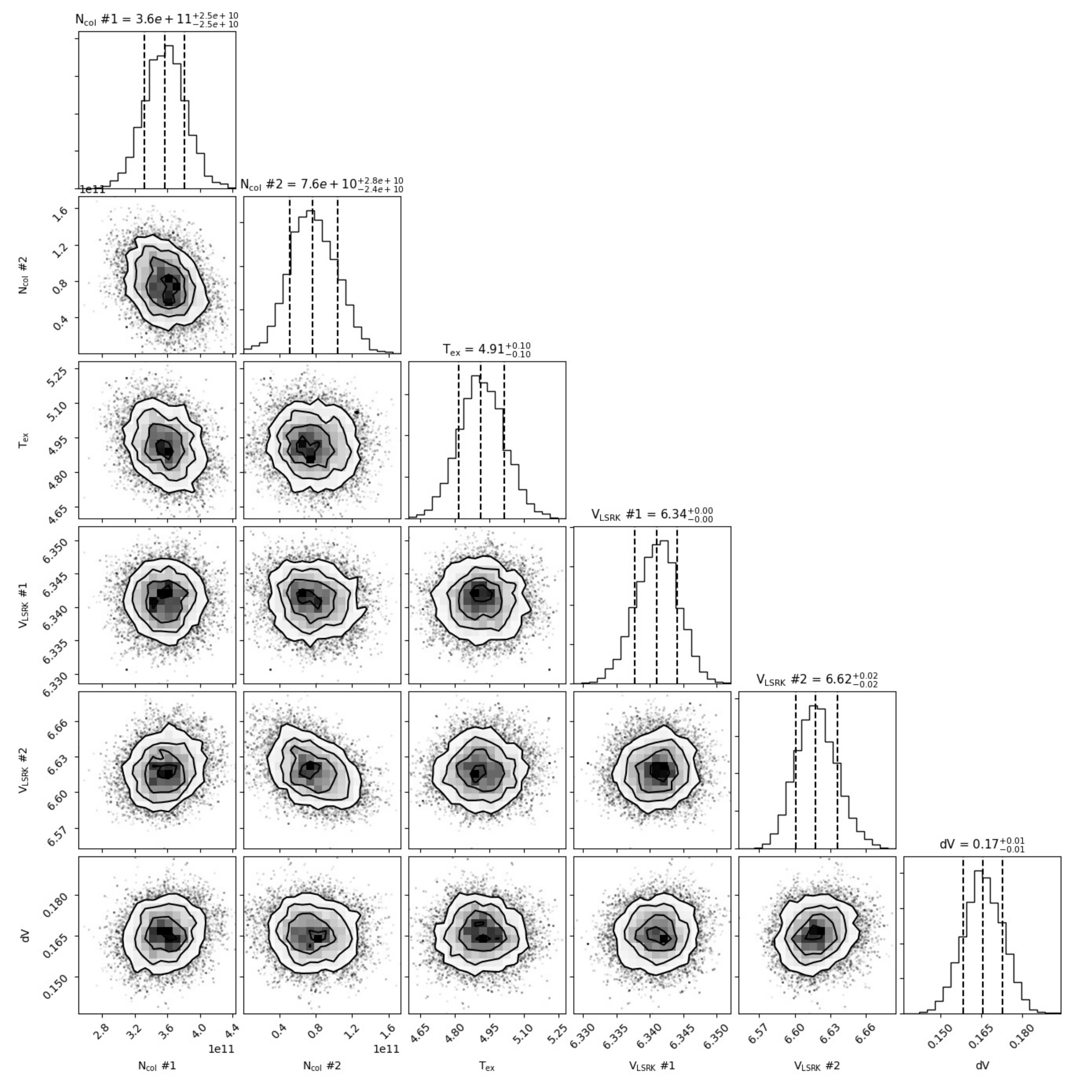}
\caption{Parameter covariances and marginalized posterior distributions for the benzonitrile MCMC fit for MC27/L1521F. 16$^{th}$, 50$^{th}$, and 84$^{th}$ confidence intervals (corresponding to $\pm$1 sigma for a Gaussian posterior distribution) are shown as vertical lines. }
\label{mc27_bn_triangle}
\end{figure}

The resulting best-fit parameters for the MCMC analysis of benzonitrile in MC27/L1512F are given in \sitablename~\ref{mc27_bn_results}. Detected individual lines are shown in \sifigurename~\ref{mc27_bn_spectra}, while the stacked spectrum and matched filter results are shown in Figure~\ref{fig:bn_stack}d. A corner plot of the parameter covariances for the benzonitrile MCMC fit is shown in \sifigurename~\ref{mc27_bn_triangle}.  Unlike the Serpens sources, the degeneracy between the column density and the excitation temperature was less pronounced in the MCMC fit. For the primary component C1, the column density had a comparitively lower uncertainties ($\sim$7\%) and the excitation temperature was similarly low ($\sim$2\%) and even lower than the uncertainity of $\Delta V$ ($\sim$4\%). Instead, the primary source of error was the uncertainties in the weaker secondary component, C2. The $v_{lsr}$ ($\sim$0.05-0.3\%) was still much component of the uncertainties than these other factors.

\begin{table}[!h]
\centering
\caption{Benzonitrile best-fit parameters for MC27/L1521F from MCMC analysis}
\begin{tabular}{c c c c c c}
\toprule
\multirow{2}{*}{Component}&	$v_{lsr}$					&	Size					&	\multicolumn{1}{c}{$N_T^\dagger$}					&	$T_{ex}$							&	$\Delta V$		\\
			&	(km s$^{-1}$)				&	($^{\prime\prime}$)		&	\multicolumn{1}{c}{(10$^{11}$ cm$^{-2}$)}		&	(K)								&	(km s$^{-1}$)	\\
\midrule
\hspace{0.1em}\vspace{-0.5em}\\
C1	&	$6.341^{+0.003}_{-0.003}$	&	[$400$]	&	$3.57^{+0.25}_{-0.25}$	&	\multirow{3}{*}{$4.9^{+0.1}_{-0.1}$}	&	\multirow{3}{*}{$0.165^{+0.007}_{-0.007}$}\\
\hspace{0.1em}\vspace{-0.5em}\\
C2	&	$6.618^{+0.018}_{-0.017}$	&	[$400$]	&	$0.76^{+0.28}_{-0.24}$	&		&	\\
\hspace{0.1em}\vspace{-0.5em}\\
\midrule
$N_T$ (Total)$^{\dagger\dagger}$	&	 \multicolumn{5}{c}{$4.33^{+0.37}_{-0.35}\times 10^{11}$~cm$^{-2}$}\\
\bottomrule
\end{tabular}

\begin{minipage}{0.75\textwidth}
	\footnotesize
	\textbf{Note} -- The quoted uncertainties represent the 16$^{th}$ and 84$^{th}$ percentile ($1\sigma$ for a Gaussian distribution) uncertainties. \\
	$^\dagger$See \sifigurename~\ref{mc27_bn_triangle} for a covariance plot, and  Loomis et al.\cite{Loomis:2020aa} for a detailed explanation of the methods used to constrain these quantities and derive the uncertainties.\\
	$^{\dagger\dagger}$Uncertainties derived by adding the uncertainties of the individual components in quadrature.
\end{minipage}

\label{mc27_bn_results}

\end{table}

\clearpage
\newpage
\subsection*{TMC-1}
\begin{figure*}
\centering
\includegraphics[width=\textwidth]{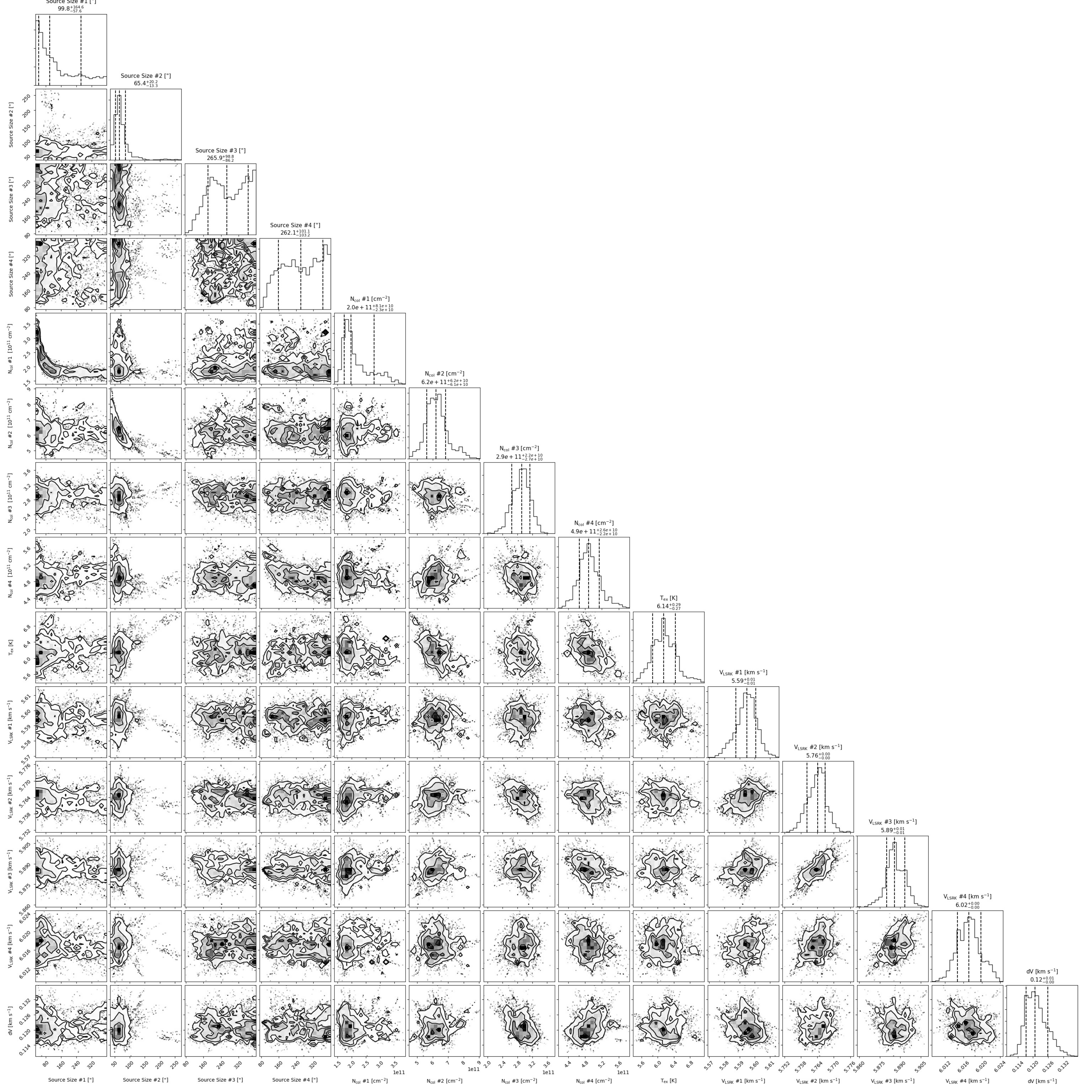}
\caption{Parameter covariances and marginalized posterior distributions for the benzonitrile MCMC fit toward TMC-1. 16$^{th}$, 50$^{th}$, and 84$^{th}$ confidence intervals (corresponding to $\pm$1 sigma for a Gaussian posterior distribution) are shown as vertical lines. Source sizes for components 3 and 4 are consistent with a source that fills the GBT beam (i.e.~they are unconstrained on the upper bound).}
\label{tmc1_bn_triangle}
\end{figure*}

The results of our MCMC fit to the dataset for $c$-\ce{C6H5CN} toward TMC-1 are provided below, and are substantially more robust than those achieved in the initial detection from \cite{McGuire:2018it}. The best-fit parameters are given in \sitablename~\ref{tmc1_bn_results}, the associated corner plot in \sifigurename~\ref{tmc1_bn_triangle}, and individually detected lines are shown and discussed in greater detail in Ref~\cite{Loomis:2020aa} and \cite{McGuire:2020bb}.

\begin{table*}[!hb]
\centering
\caption{\ce{Benzonitrile} best-fit parameters for TMC-1 from MCMC analysis}
\begin{tabular}{c c c c c c}
\toprule
\multirow{2}{*}{Component}&	$v_{lsr}$					&	Size					&	\multicolumn{1}{c}{$N_T^\dagger$}					&	$T_{ex}$							&	$\Delta V$		\\
			&	(km s$^{-1}$)				&	($^{\prime\prime}$)		&	\multicolumn{1}{c}{(10$^{11}$ cm$^{-2}$)}		&	(K)								&	(km s$^{-1}$)	\\
\midrule
\hspace{0.1em}\vspace{-0.5em}\\
C1	&	$5.595^{+0.006}_{-0.007}$	&	$99^{+164}_{-57}$	&	$1.98^{+0.81}_{-0.23}$	&	\multirow{6}{*}{$6.1^{+0.3}_{-0.3}$}	&	\multirow{6}{*}{$0.121^{+0.005}_{-0.004}$}\\
\hspace{0.1em}\vspace{-0.5em}\\
C2	&	$5.764^{+0.003}_{-0.004}$	&	$65^{+20}_{-13}$	&	$6.22^{+0.62}_{-0.61}$	&		&	\\
\hspace{0.1em}\vspace{-0.5em}\\
C3	&	$5.886^{+0.007}_{-0.006}$	&	$265^{+98}_{-86}$	&	$2.92^{+0.22}_{-0.27}$	&		&	\\
\hspace{0.1em}\vspace{-0.5em}\\
C4	&	$6.017^{+0.003}_{-0.002}$	&	$262^{+101}_{-103}$	&	$4.88^{+0.26}_{-0.22}$	&		&	\\
\hspace{0.1em}\vspace{-0.5em}\\
\midrule
$N_T$ (Total)$^{\dagger\dagger}$	&	 \multicolumn{5}{c}{$1.60^{+0.11}_{-0.07}\times 10^{12}$~cm$^{-2}$}\\
\bottomrule
\end{tabular}

\begin{minipage}{0.75\textwidth}
	\footnotesize
	\textbf{Note} -- The quoted uncertainties represent the 16$^{th}$ and 84$^{th}$ percentile ($1\sigma$ for a Gaussian distribution) uncertainties.\\
	$^\dagger$Column density values are highly covariant with the derived source sizes. The marginalized uncertainties on the column densities are therefore dominated by the largely unconstrained nature of the source sizes, and not by the signal-to-noise of the observations. See \sifigurename~\ref{tmc1_bn_triangle} for a covariance plot, and Ref.~\cite{Loomis:2020aa} for a detailed explanation of the methods used to constrain these quantities and derive uncertainties.\\
	$^{\dagger\dagger}$Uncertainties derived by adding the uncertainties of the individual components in quadrature.
\end{minipage}

\label{tmc1_bn_results}

\end{table*}

\clearpage
\newpage
\section*{Additions to \textsc{nautilus} chemical model and \ce{C6H6}/\ce{C6H5CN} formation}
Building off of the work in \cite{McGuire:2018it}, we have further updated to the chemical network to better describe the formation of benzene, and subsequently benzonitrile. Due to the significant underproduction of the initial model from \cite{McGuire:2018it} to reproduce the newly derived column densities with the methods described in \cite{Loomis:2020aa}, we significantly expanded the reaction network to include additional production and destruction pathways, as described in \sitablenames~\ref{tab:newgasnetwork} and \ref{tab:newsolidnetwork}. This includes pathways for \ce{C6H6}, as well as some of potentially important precursors \ce{C6H7+}, 1,3-butadiene, and propene. Also, the rate coefficient for \ce{CN + C6H6} is increased from 3$\times$10$^{-10}$ cm$^3$ s$^{-1}$ \cite{Woon:2006ce} to 5$\times$10$^{-10}$ cm$^3$ s$^{-1}$ \cite{Cooke:2020we}. As such, while the network is by no means complete, it does at minimum extensively describe a more accurate description of the current capabilities of kinetic chemical models to reproduce aromatic abundances without additional processes.

\begin{table}[!h]
\centering
\caption{Baseline physical conditions and parameters for \textsc{nautilus} models. }
\begin{tabular}{c c}
\toprule
$T_{\text{gas}}$    & 10 K\\
$T_{\text{grain}}$  & 10 K\\
$A_V$               & 10 \\
$n_{\text{H}_2}$    & 2$\times$10$^4$ cm$^{-3}$\\
$\zeta_{\text{CR}}$                & 1.3$\times$10$^{-17}$ s$^{-1}$\\
\bottomrule
\end{tabular}

\begin{minipage}{0.75\textwidth}
	\footnotesize
	\textbf{Note} -- Physical conditions described here also hold for all chemical models shown in GOTHAM or ARKHAM papers, except when explicitly described. 
\end{minipage}

\label{tab:model_parameters}

\end{table}

\begin{table}[!h]
\centering
\caption{Baseline initial elemental for \textsc{nautilus} models. }
\begin{tabular}{c c c}
\toprule
Species & Abundance & Reference\\
\midrule
\ce{H2} & 0.5\\
H       & 5.00$\times$10$^{-5}$ & \\
He      & 9.00$\times$10$^{-2}$ & a\\
C       & 1.70$\times$10$^{-4}$ & b\\
N       & 6.20$\times$10$^{-5}$ & b\\
O       & 1.54$\times$10$^{-4}$ & $\star$\\
F       & 6.68$\times$10$^{-9}$ & c\\
Na      & 2.00$\times$10$^{-9}$ & d\\
Mg      & 7.00$\times$10$^{-9}$ & d\\
Si      & 8.00$\times$10$^{-9}$ & d\\
P       & 2.00$\times$10$^{-10}$ & d\\
S       & 8.00$\times$10$^{-8}$ & d\\
Cl      & 1.00$\times$10$^{-9}$ & d\\
Fe      & 3.00$\times$10$^{-9}$ & d\\
\bottomrule
\end{tabular}

\begin{minipage}{0.75\textwidth}
	\footnotesize
	\textbf{Note} -- ($\star$) C/O ratio was constrained by the model that best fit the measured abundances of the detected cyanopolyyne towards TMC-1 in the GOTHAM survey. Otherwise, adopted values utilized by \cite{hincelin_oxygen_2011}.\\ 
	\textbf{References} -- (a) Wakelam \& Herbst \cite{wakelam:2008}; (b) Jenkins \cite{Jenkins:2009ke}; (c)  Neufeld et al. \cite{Neufeld:2005lk}; (d) Graedel et al. \cite{Graedel:1982}
\end{minipage}

\label{tab:model_elements}

\end{table}

\begin{landscape}

\small
    \begin{center}
\begin{longtable}{r  l c c c c c c c}
\caption{Gas-Phase Benzene-Relevant Additions to Chemical Network from \cite{McGuire:2018it}} \label{tab:newgasnetwork} \\
\hline 
\multicolumn{2}{c}{Reaction(s)} & 
\multicolumn{1}{c}{$\alpha$} & 
\multicolumn{1}{c}{$\beta$} & 
\multicolumn{1}{c}{$\gamma$} & 
\multicolumn{1}{c}{$T_{\text{min}}$} & 
\multicolumn{1}{c}{$T_{\text{max}}$} & 
\multicolumn{1}{c}{Reaction Type} &
\multicolumn{1}{c}{Reference} \\ 
\hline
\endfirsthead

\multicolumn{3}{c}%
{{\bfseries \sitablename\ \thetable{} -- continued from previous page}} \\
\hline 
\multicolumn{2}{c}{Reaction(s)} & 
\multicolumn{1}{c}{$\alpha$} & 
\multicolumn{1}{c}{$\beta$} & 
\multicolumn{1}{c}{$\gamma$} & 
\multicolumn{1}{c}{$T_{\text{min}}$} & 
\multicolumn{1}{c}{$T_{\text{max}}$} & 
\multicolumn{1}{c}{Reaction Type} &
\multicolumn{1}{c}{Reference} \\ 
\hline
\endhead

\endfoot

\ce{C6H6}  +     h$\nu_\mathrm{internal}$        & \hspace{-3mm}$\longrightarrow$         \ce{H}     +    \ce{C6H}     &      2.925$\times10^{3}$   & 0 & 0 & -  & - & 1 & a  \\
                            & \hspace{-3mm}$\longrightarrow$         \ce{H2}   +      \ce{C6H4}     &    6.210$\times10^{1}$   & 0 & 0 & -  & - & 1 & a  \\
\ce{C6H6}  +     h$\nu_\mathrm{ISRF}$   & \hspace{-3mm}$\longrightarrow$         \ce{H}     +    \ce{C6H}     &      1.950$\times10^{-11}$ & 0 & 1.7 & -  & - & 2 & a  \\
                            & \hspace{-3mm}$\longrightarrow$         \ce{H2}   +      \ce{C6H4}     &    4.140$\times10^{-13}$ & 0 & 1.7 & -  & - & 2 & a  \\
\hline
\ce{CN}      +   \ce{C6H6}     & \hspace{-3mm}$\longrightarrow$         \ce{C6H5CN}  +  \ce{H}     			& 4.500$\times10^{-10}$ & 0 & 0 &  - &  - & 3 & a,b \\
\ce{C6H5CN}  +   \ce{H3+}      & \hspace{-3mm}$\longrightarrow$         \ce{C6H5+}   +   \ce{HCN}    +    \ce{H2}   & 1 & 6.180$\times10^{-9}$ & 63.91 &  - &   - & 4 &  a  \\
\ce{C6H5CN}  +   \ce{H+}       & \hspace{-3mm}$\longrightarrow$         \ce{C6H5+}   +   \ce{HCN}     	        & 1 & 1.059$\times10^{-8}$ & 63.91 &  - &    & 4 &  a  \\
\ce{C6H5CN}  +   \ce{He+}      & \hspace{-3mm}$\longrightarrow$         \ce{C6H5+}   +  \ce{CN}     +   \ce{He}     & 1 & 7.529$\times10^{-9}$ & 63.91 &  -  &  - & 4 &  a  \\
\ce{C6H5CN}  +   \ce{C+}       & \hspace{-3mm}$\longrightarrow$         \ce{C6H5+}   +   \ce{CCN}     		& 1 & 3.234$\times10^{-9}$ & 63.91 &  - &  - & 4 &  a  \\
\ce{C6H5CN}  +   \ce{HCO+}     & \hspace{-3mm}$\longrightarrow$         \ce{C6H5+} +     \ce{HCO}  +     \ce{CN}    & 1 & 2.245$\times10^{-9}$ & 63.91 &  -  &  - & 4 &  a  \\
\hline
\ce{CH}       +  \ce{C2H6}     & \hspace{-3mm}$\longrightarrow$         \ce{CH3CHCH2} + \ce{H}     			& 5.000$\times10^{-12}$ & $-6.480\times10^{-1}$ & $4.360\times10^1$ &     10  &  800 & 3 &  c  \\
                               & \hspace{-3mm}$\longrightarrow$         \ce{CH3}      +  \ce{C2H4}     		& 2.800$\times10^{-11}$ & $-6.480\times10^{-1}$ & $4.360\times10^1$ &     10  &  800 & 3 &  c  \\
\ce{H3+}      +  \ce{CH3CHCH2} & \hspace{-3mm}$\longrightarrow$         \ce{C2H3+}   +   \ce{CH4}     +   \ce{H2}   & 3.000$\times10^{-1}$ & $3.500\times10^{-9}$ & $5.100\times10^{-1}$ &     10  &  800 & 3 &  d  \\
                               & \hspace{-3mm}$\longrightarrow$         \ce{C3H5+}   +   \ce{H2}      +   \ce{H2}   & 3.000$\times10^{-1}$ & $3.500\times10^{-9}$ & $5.100\times10^{-1}$ &     10  &  800 & 3 &  d  \\
\ce{C+}       +  \ce{CH3CHCH2} & \hspace{-3mm}$\longrightarrow$         \ce{C2H3+}   +   \ce{C2H3}     		& 6.000$\times10^{-10}$ & $-5.000\times10^{-1}$ & 0 &     10 &   800 & 3 &  d  \\
                               & \hspace{-3mm}$\longrightarrow$         \ce{C3H5+}   +   \ce{CH}     		& 4.000$\times10^{-10}$ & $-5.000\times10^{-1}$ & 0 &     10 &   800 & 3 &  d  \\
                               & \hspace{-3mm}$\longrightarrow$         \ce{l-C3H3+} +   \ce{CH3}     		& 3.000$\times10^{-10}$ & $-5.000\times10^{-1}$ & 0 &     10 &   800 & 3 &  d  \\
                               & \hspace{-3mm}$\longrightarrow$         \ce{C2H2+}   +   \ce{C2H4}     		& 3.000$\times10^{-10}$ & $-5.000\times10^{-1}$ & 0 &     10 &   800 & 3 &  d  \\
                               & \hspace{-3mm}$\longrightarrow$         \ce{C4H3+}   +   \ce{H2}    +    \ce{H}     & 2.000$\times10^{-10}$ & $-5.000\times10^{-1}$ & 0 &     10   & 800 & 3 &  d  \\
\hline
%
%
\ce{CCH}     +   \ce{CH2CHCHCH2}    & \hspace{-3mm}$\longrightarrow$         \ce{H}   +      \ce{C6H6}     &        1.000$\times10^{-10}$ & 0  &0 &  - &  -&  3 &  e   \\
\hline
\ce{C2H6}   +   h$\nu_\mathrm{internal}$      & \hspace{-3mm}$\longrightarrow$         \ce{H2}      +   \ce{C2H4}     &        1.880$\times10^{3}$ & 0 & 0 &  -  & - & 1 &  f  \\
                         & \hspace{-3mm}$\longrightarrow$         \ce{C2H6+}   +   \ce{e-}     &          3.890$\times10^{2}$ & 0 & 0 &  -  & - & 1 &  f  \\
\hline
\ce{CN}    +     \ce{C2H2}   & \hspace{-3mm}$\longrightarrow$         \ce{H}   +      \ce{HC3N}     &       5.260$\times10^{-9}$ & -0.52 & 20.0 &  10  &  300 & 3 &  g  \\
\hline
\ce{H2}       +  \ce{C6H5}   & \hspace{-3mm}$\longrightarrow$         \ce{H}      +   \ce{C6H6}     &          9.910$\times10^{-14}$ & 2.430 & 3160 &     50  &  200  & 3 & g \\
\ce{H}        +  \ce{C6H6}     & \hspace{-3mm}$\longrightarrow$         \ce{H2}  +       \ce{C6H5}     &       4.150$\times10^{-10}$ & 0 & 8050  &     50 &   200 & 3 & g \\
\ce{OH}       +  \ce{C6H6}     & \hspace{-3mm}$\longrightarrow$         \ce{H2O} +       \ce{C6H5}     &       2.800$\times10^{-11}$ & 0 & 2302 &    298 &  1500 & 3 & h  \\
\hline
\ce{C2H4}     +  \ce{C4H3}   & \hspace{-3mm}$\longrightarrow$         \ce{H}     +    \ce{C6H6}     &         1.500$\times10^{-10}$ & 0 & 0 &     50  &  200 & 3 & g \\
\hline
\ce{CH3CHCH2} +  \ce{C4H3}    & \hspace{-3mm}$\longrightarrow$         \ce{CH3}  +      \ce{C6H6}     &       1.200$\times10^{-12}$ & 0 & 2520  &     50  &  200 & 3 & g \\
\ce{CCH}      +  \ce{CH3CHCH2} & \hspace{-3mm}$\longrightarrow$         \ce{CH}  +       \ce{CH2CHCHCH2}   &  2.000$\times10^{-11}$ & 0 & 0 &     50 &   200 & 3 & g \\
\ce{C2H3}     +  \ce{C2H4}    & \hspace{-3mm}$\longrightarrow$         \ce{H}    +     \ce{CH2CHCHCH2}     &  8.300$\times10^{-13}$ & 0 & 3680 &    50 &   200 & 3 & g \\
\ce{C2H3}     +  \ce{CH3CHCH2} & \hspace{-3mm}$\longrightarrow$         \ce{CH3} +       \ce{CH2CHCHCH2}   &  1.200$\times10^{-12}$ & 0 & 2520 &     50  &  200 & 3 & g \\
\ce{C}        +  \ce{CH2CHCHCH2}   & \hspace{-3mm}$\longrightarrow$     \ce{C2H3} +    \ce{CH2CCH}     &      1.100$\times10^{-9}$  & 0 & 0 &     50  &  200 & 3 & c \\
\ce{H}        +  \ce{CH3CHCH2} & \hspace{-3mm}$\longrightarrow$         \ce{CH3} +       \ce{C2H4}     &      1.200$\times10^{-11}$ & 0 & 655  &     50  &  200 & 3 & g \\
\ce{CN}       +  \ce{CH3CHCH2} & \hspace{-3mm}$\longrightarrow$         \ce{C2H3}  +   \ce{CH3CN}  & 1.730$\times10^{-10}$ & 0 & -102 & 50 &   200 & 3 &  g \\
\ce{C+}       +  \ce{CH3CHCH2}  & \hspace{-3mm}$\longrightarrow$         \ce{c-C3H3+} +    \ce{CH3}     &     1.500$\times10^{-10}$ & -0.5 & 0 &     10  &  800 & 3 & c \\
\ce{C}        +  \ce{CH3CHCH2} & \hspace{-3mm}$\longrightarrow$         \ce{CH3}      +  \ce{CH2CCH}     &    2.000$\times10^{-10}$ & 0 & 0 &     10  &  800 & 3 & c  \\
\ce{O}        +  \ce{CH3CHCH2}  & \hspace{-3mm}$\longrightarrow$         \ce{C2H5}    +    \ce{HCO}     &     3.600$\times10^{-12}$ & 0 & 0 &     10  &  800 & 3 & c \\
\ce{CH}       +  \ce{C2H6}   & \hspace{-3mm}$\longrightarrow$         \ce{H}     +    \ce{CH3CHCH2}     &     5.000$\times10^{-12}$ & -0.648 & 43.6 &     10 &   800 & 3 & c  \\
\ce{C2H5}     +  \ce{CH2}     & \hspace{-3mm}$\longrightarrow$         \ce{H}    +     \ce{CH3CHCH2}     &    1.500$\times10^{-11}$ & 0 & 0 &     50  &  200 & 3 & g \\
\hline
\end{longtable}

\begin{minipage}{\textwidth}
	\footnotesize
	\textbf{Note} --
	Formulae of type 1 and 2 are $k = \alpha \zeta$ and $k = \alpha e^{-\gamma A_\nu}$, where $k$ is in $\mathrm{s^{-1}}$, and formulae of type 3 and 4 are $k(T) = \alpha \left(T/300\right)^\beta e^{-\gamma/T}$ and $k(T) = \alpha \beta \left(0.62 + 0.4767 \gamma \left(300/T\right)^{0.5}\right)$, where $k$ is in $\mathrm{cm^3\,s^{-1}}$ and $T$ is in $\mathrm{K}$, respectively.\\
	(a) Dissociation pathways and branching fractions taken from Kislov et al. \cite{Kislov:2004} using the 154 nm values, representative of the 80-150 nm internal UV photon\\
	(b) Cooke et al. \cite{Cooke:2020we} \\
    (c) Loison et al \cite{Loison:2017}
	(d) Hickson et al. \cite{Hickson:2016}\\
	(e) Jones et al. \cite{Jones:2011yc} \\
	(f) Balucani, Ceccarelli, and Taquet \cite{Balucani:2015gj} \\
	(g) Hebrard et al \cite{Hebrard:2009}\\
	(h) Baulch et al \cite{Baulch:2005} \\ 
\end{minipage}

\end{center}

\clearpage 

\small
\begin{center}
\begin{longtable}{r  l c c c }
\caption{Solid-Phase Benzene-Relevant Additions to Chemical Network from \cite{McGuire:2018it}} \label{tab:newsolidnetwork} \\
\hline 
\multicolumn{2}{c}{Reaction(s)} & 
\multicolumn{1}{c}{$\alpha$} & 
\multicolumn{1}{c}{$\beta$} & 
\multicolumn{1}{c}{$\gamma$}   \\ 
\hline
\endfirsthead

\endhead

\endfoot
\ce{H}     +    \ce{C6H5} & \hspace{-3mm}$\longrightarrow$ \ce{C6H6} &  1& 0 & 0 \\
\ce{CN}    +    \ce{C6H5} & \hspace{-3mm}$\longrightarrow$ \ce{C6H5CN} & 1& 0 & 0  \\
\ce{H}     +    \ce{C6H5CN} & \hspace{-3mm}$\longrightarrow$ \ce{C6H5}   +   \ce{HCN} &  1& 0 & 4455 \\
\ce{CH}      +  \ce{C2H6} & \hspace{-3mm}$\longrightarrow$ \ce{CH3}   +     \ce{C2H4}      & 0.75 & 0 &  0 \\
                           & \hspace{-3mm}$\longrightarrow$ \ce{H}       +   \ce{CH3CHCH2}   & 0.25 & 0 &  0 \\ 
\ce{CH}  +  \ce{CH3CHCH2} & \hspace{-3mm}$\longrightarrow$  \ce{CH3} +       \ce{CH3CCH}     & 0.1  & 0 &  0  \\
                           & \hspace{-3mm}$\longrightarrow$ \ce{H}     +     \ce{CH2CHCHCH2}  & 0.9 & 0 &  0 \\
\hline
\end{longtable}

\begin{minipage}{\textwidth}
	\footnotesize
	\textbf{Note} -- Only bimolecular reactions are listed here, although adsorption, desporption, and surface/mantle exchange are included in the network. 
\end{minipage}
\end{center}

\end{landscape}

\section*{Additional comparison of \textsc{nautilus} chemical model for \ce{C6H5CN} and \ce{HC9N}}
\begin{figure}
    \centering
    \includegraphics[width=0.49\textwidth]{C6H5CN_grid_tgas_nh2_20x20.pdf}
    \includegraphics[width=0.49\textwidth]{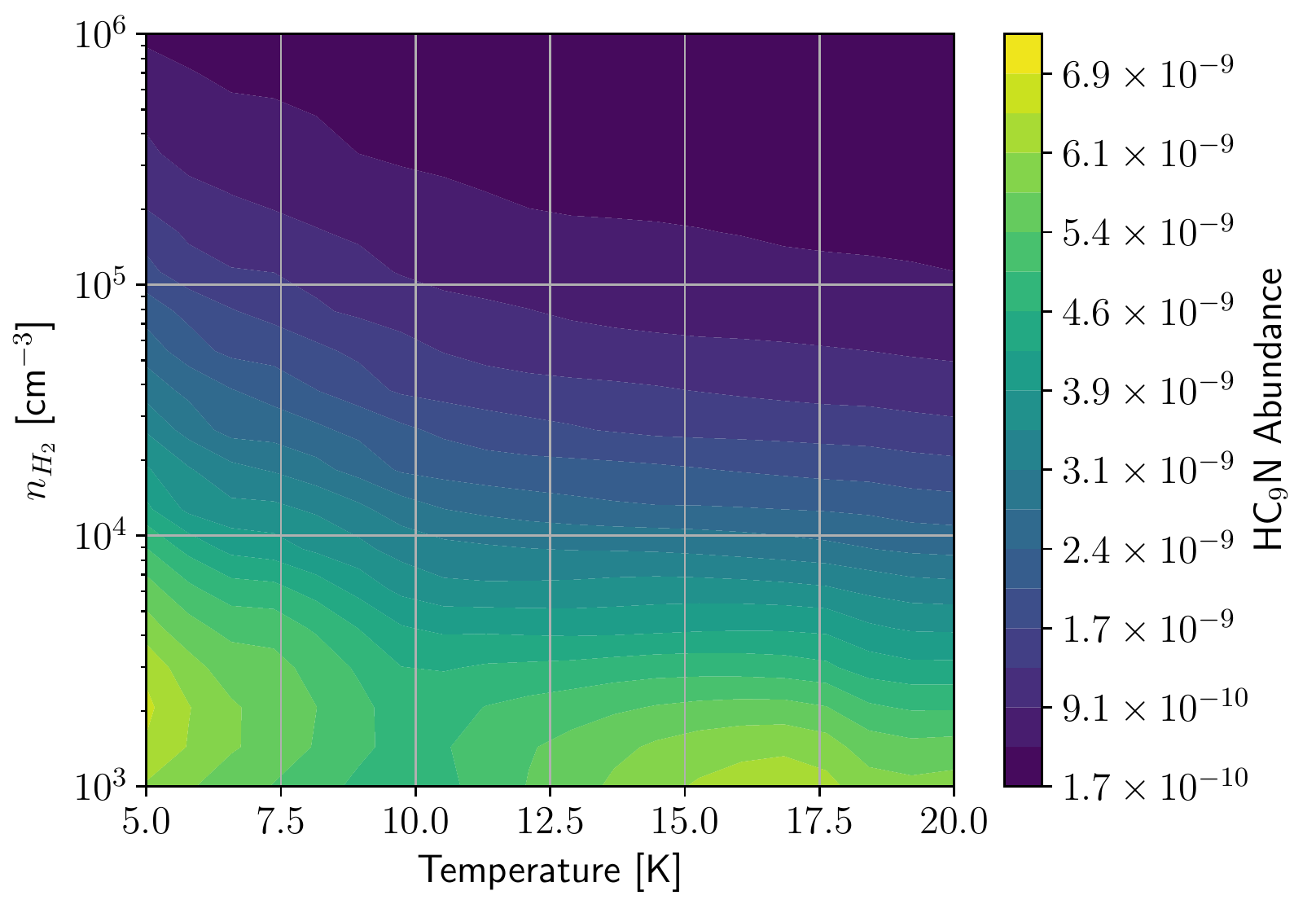}
    \includegraphics[width=0.49\textwidth]{C6H5CN_grid_tgas_O_20x20.pdf}
    \includegraphics[width=0.49\textwidth]{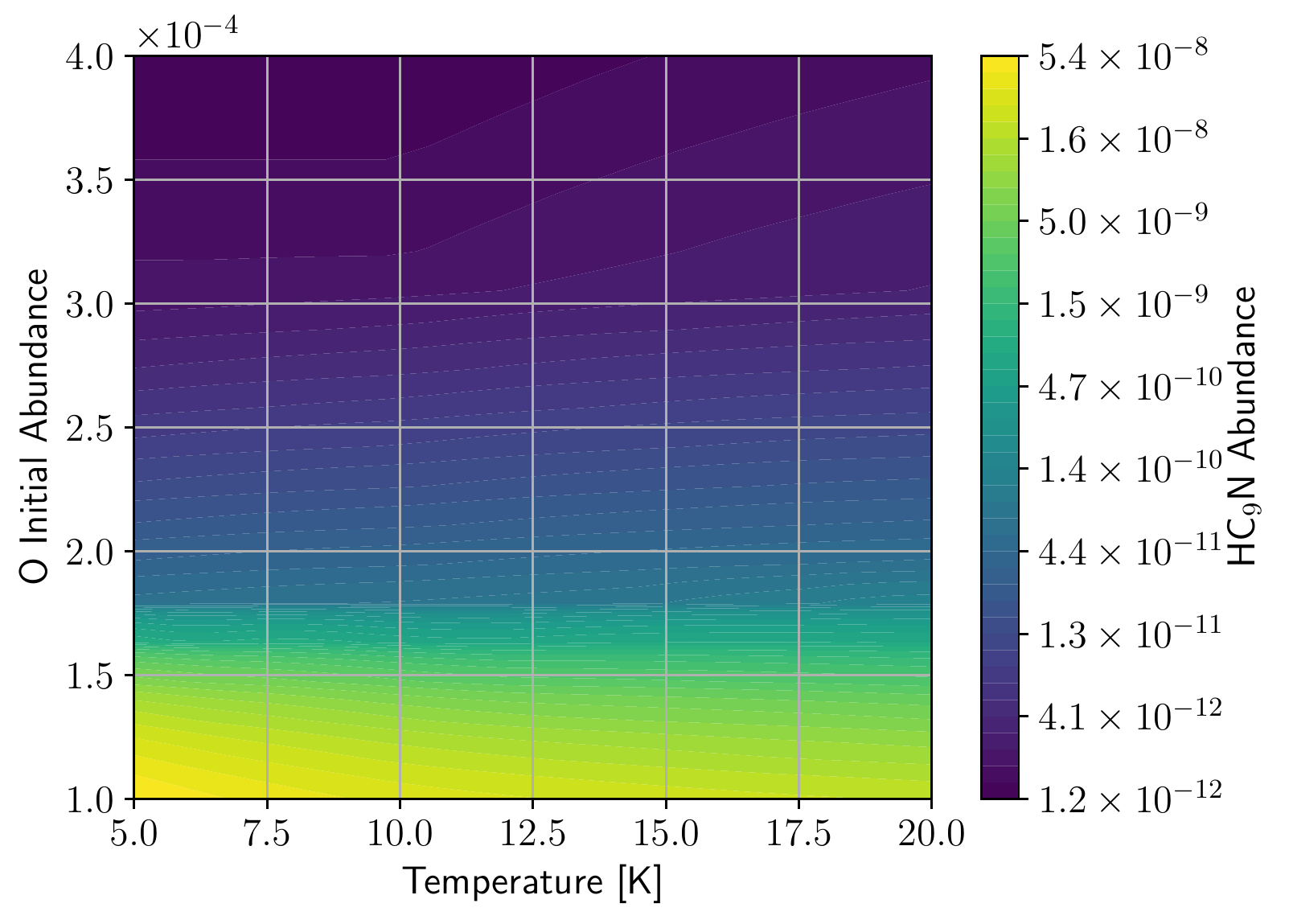}
    \includegraphics[width=0.49\textwidth]{C6H5CN_grid_nh2_O_20x20.pdf}
    \includegraphics[width=0.49\textwidth]{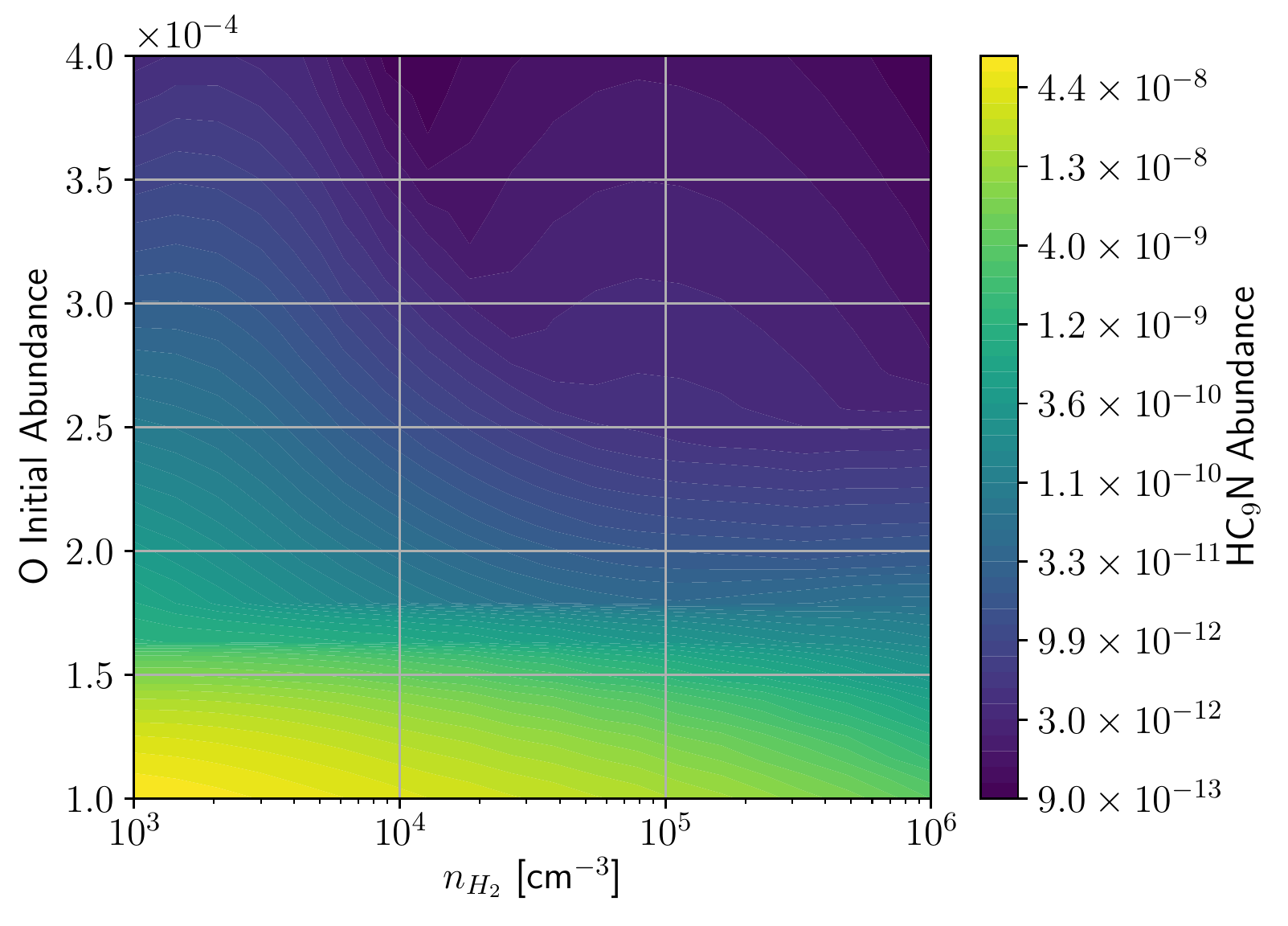}
    \caption{Simulated abundances of \ce{C6H5CN} (\emph{Left}) and \ce{HC9N} (\emph{Right}) from \textsc{nautilus} chemical models over a range of gas and grain temperatures, gas densities, and initial elemental oxygen abundances. }
    \label{fig:grids_hc9n}
\end{figure}

\sifigurename~\ref{fig:grids_hc9n} shows the simulated abundances of \ce{C6H5CN} and \ce{HC9N} with \textsc{nautilus} over the same range of gas and grain temperatures, gas densities, and initial elemental oxygen abundances as Figure \ref{fig:grids}. As mentioned in the Astrochemical Modeling section of Methods, both species are strongly correlated with an increase in the C/O ratio. However, \ce{HC9N} appears less sensitive to changes in temperature than \ce{C6H5CN}. In terms of gas density, \ce{HC9N} is likely destroyed more efficiently at higher densities, \ce{C6H5CN} is most abundant within a definite density regime.  

\end{document}